\documentclass[12pt]{article}

\usepackage[left=1in,top=1in,right=1in,bottom=1in]{geometry} % Change margin size.
\usepackage{graphicx,calc} % Inserting graphics.
\usepackage[hidelinks]{hyperref} % Hides hyperlink boxes.
\usepackage[dvipsnames]{xcolor} % Coloured headings.
\usepackage{physics} % Shorthand derivatives and other maths commands.
\usepackage{amssymb} % Maths fonts such as \mathcal{R} and \mathbb{R}.
\usepackage{bbm} % gives doubble font for numbers too, \mathbbm{1}
\usepackage{enumitem} % Package for lists.
\usepackage{mathtools} % For overlaying maths characters.
\usepackage{blindtext} % For making a minipage contain blank space.
\usepackage[enableskew]{youngtab} %For Young Tableau. \yng(2,1) for empty tableau with 2 boxes in 1st row and 1 in second. \young(aab,b) for 3 boxes with a a b in 1st row and 1 box with b in 2nd row.
\usepackage{slashed} % Feynman slash notation. \slashed{A}.
\usepackage{simplewick} % For wick contractions.
\usepackage[compat=1.0.0]{tikz-feynman} % Feynman diagrams
\usepackage{mhchem} %chemical equilibrium arrows
\usepackage{stackengine} %\stackanchor[8pt]{top item}{bottom item}
\usepackage[titletoc]{appendix} % appendix

\setitemize{leftmargin=*} % Compacts list format.

\usepackage{array}
\newcolumntype{C}[1]{>{\centering\let\newline\\\arraybackslash\hspace{0pt}}m{#1}}	% new column type.

\usepackage{booktabs} %\midrule for better \hline in tables

\usepackage{setspace} %For text double spacing \doublespacing

\newcommand{\tb}{\textbf} % Shorthand for bold text
\newcommand{\ti}{\textit} % Shorthand for italic text
\newcommand{\mb}{\mathbf} % Shorthand for bold maths text
\newcommand{\mbb}{\mathbbm} % Shorthand for bold maths text
\newcommand{\p}{\partial} % Shorthand for partial sign
\newcommand{\Lie}{\mathcal{L}} % Shorthand for Lie-algebra
\newcommand{\Od}{\mathcal{O}} % Shorthand for Order symbol
\newcommand{\pink}{\textcolor{magenta}} % Shorthand for pink text

\begin{document}

\title{The Phenomenological Motivation of Axions: A Review}
\author{Drew Backhouse}
\date{}
\maketitle

\begin{center}
\includegraphics[width=0.4\linewidth]{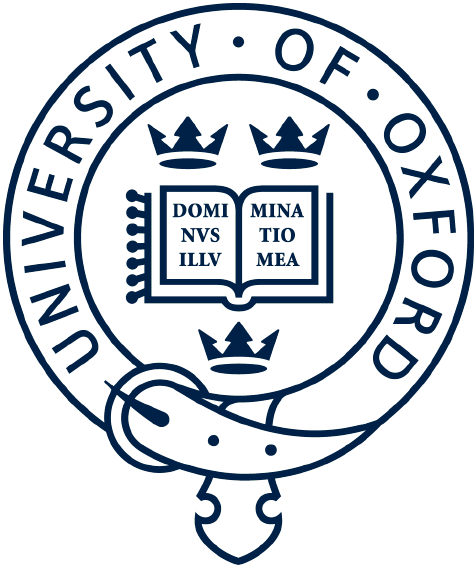}
\end{center}

\vspace{1in}

\begin{center}
\Large{Dissertation Submitted in Partial Fulfilment of the Requirements for the Degree of Master of Mathematical and Theoretical Physics\\
\vspace{0.5in}
University of Oxford\\
Trinity 2021}
\end{center}

\newpage

\begin{center}
\tb{\large{Abstract}}
\end{center}

Setting aside anthropic arguments, there is no reason for CP symmetry to be obeyed within the theory of quantum chromodynamics. However, no such violation of CP symmetry has ever been observed in a strongly interacting experiment. This is known as \ti{the strong CP problem} which, in its simplest manifestations, can be quantitatively formulated via a calculation of the pion masses and the neutron electric dipole moment. The former yields a larger mass for the neutral pion than its charged counterparts, the latter yields a far larger result than is experimentally measured, where in both cases the discrepancies are parameterised by the physical quantity $\bar{\theta}$. The strong CP problem can be solved via the inclusion of a new particle, \ti{the axion}, which dynamically sets $\bar{\theta}$ to zero, eliminating these two manifestations. Thus, experimental searches for such a particle are an active field of research. This dissertation acts as a review of the aforementioned concepts.

\newpage

\tableofcontents

\newpage

\section{Introduction}

The reason I can write this dissertation is because, for some inexplicable reason, there is significantly more matter present in the universe than antimatter and thus the whole universe does not simply annihilate into photons. The question is, why am I not just a photon propagating in an empty universe?\\ %Joe not a fan of last sentence.

In 1957, Lev Landau proposed that CP-symmetry is the true symmetry between matter and antimatter \cite{Landau}. CP-symmetry is the invariance of a system under two successive transformations: Charge conjugation (C) and Parity (P). These successive transformations send all particles to their antiparticles and perform a coordinate inversion. If CP-symmetry were obeyed, then equal amounts of matter and antimatter would have been created during the big bang. For anything to exist at all, these must have been separated into totally non-interacting clusters before the universe dropped below about 500 billion Kelvin, otherwise, the matter and antimatter would have mutually annihilated into photons \cite{Scott Dodelson}. However, at this time the longest distances in causal contact were about 100 km, a billionth the size of any independent astronomical clusters or galaxies we observe today. Thus, simply from causality, the fact we are here today suggests there is not a perfect symmetry between matter and antimatter, undeniable proof that CP-symmetry is allowed to be broken.\\

Our current mathematical formulation of quantum chromodynamics (QCD) allows for CP-violating terms to be added to the Lagrangian, a prospect we are now totally comfortable with. However, we have never experimentally observed CP-violation in QCD. An example of this is the magnitude of the neutron electric dipole moment (eDM), which (when including the CP-violating terms in the Lagrangian) has an experimental upper limit far smaller than QCD predicts. This can be solved by setting the parameters of the CP-violating terms to zero; this requirement of `fine tuning' our theory is known as the \ti{strong CP problem}.\\

The strong CP problem is considered an unsolved problem in physics. There are many proposed solutions, one of which is the existence of a new particle called the \ti{axion}. In this Dissertation, we quantitatively formulate the strong CP problem, explaining how the axion solves it, before generalising to the more abstract \ti{axion-like particle} and discussing various methods of experimentally probing them.\\

We will assume knowledge of quantum field theory at the level of a masters course. Appendix \ref{A} contains some reference formulae which we will make use of. If the reader is in search of a detailed discussion of the basic concepts, I find the book by Mark Srednicki \cite{QFT} to be a very good read, however, be warned he uses the mostly positive metric convention whereas we shall adopt the more commonly used mostly negative metric convention ($g=(+---)$). The book by Michael E. Peskin and Daniel V. Schroder \cite{PS} is another standard text and has a very nice appendix containing all the tools you could ever need (and uses our metric convention).

\newpage

\section{The Classical Solution To The Strong CP Problem}

The existence of a CP-violating strong interaction would result in a predicted neutron electric dipole moment (eDM) of $d_n\sim10^{-16}e$ cm, while the current experimental upper bound is roughly one-billionth the size \cite{Baker}. Thus, in one of its simplest manifestations, the strong CP problem is the dilemma that the neutron eDM is measured to be far smaller than we calculate it to be.\\

\begin{figure}[h!]
\centering
\includegraphics[width=0.4\linewidth]{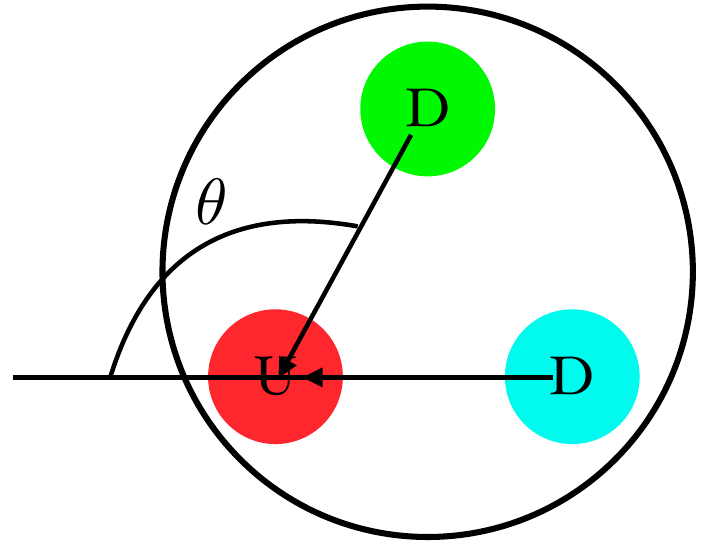}
\caption{A classical picture of the neutron with angle $\theta$ between the up and down quark \cite{TASI}.}
\label{Neutron}
\end{figure}

To begin finding a solution to this problem, we refer the reader to the classical picture of the neutron, as depicted in figure \ref{Neutron}. It is the grouping of three equally spaced quarks: one up quark of charge $\frac{2}{3}e$ and two down quarks of charge $-\frac{1}{3}e$, where $e$ is the electronic charge. Calculating the eDM of the neutron using the classical formula \begin{equation}\label{classical eDM 1} \vec{d}=\sum_i q_i \vec{r_i}\end{equation} using a neutron size $r_{n} \sim \frac{1}{m_{\pi}}$, yields a result: \begin{equation}\label{classical eDM 2} d_{n} \approx 10^{-13} \sqrt{1-\cos \theta} e \mathrm{~cm}.\end{equation} For any reasonable value of $\theta$, corresponding to equally spaced quarks, our calculated eDM is of order $d_n\sim 10^{-13}$.\\

An upper limit for the neutron eDM can be experimentally obtained \cite{Baker}, with the current best measurement reading: \begin{equation}\label{eDM} \left|d_{n}\right| \leq 10^{-26} e \mathrm{~cm}\;(90\% \text{ C.L.})\end{equation} and thus the strong CP problem is bestowed upon us, this measurement yields a far smaller value than our calculation.\\

\begin{figure}[h!]
\centering
\includegraphics[width=0.75\linewidth]{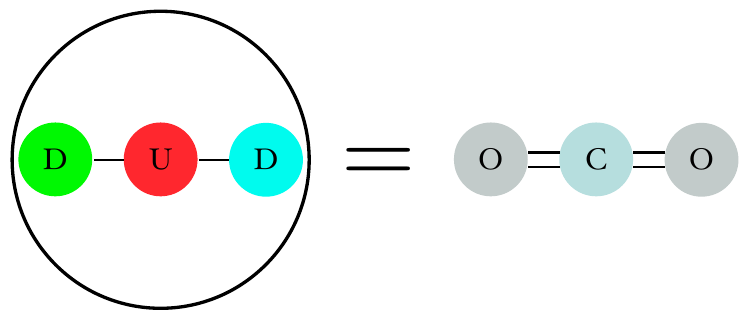}
\caption{The classical axion solution to the strong CP problem \cite{TASI}.}
\label{Axion}
\end{figure}

There is, however, a very simple solution. Taking the angle between the quarks ($\theta$) close to zero yields a neutron comprising of aligned quarks, as seen in figure \ref{Axion}. Although different to our usual picture of the neutron, this configuration is not quite as exotic as one might think. Given the angle between the up and down quark is dynamical, the quarks will stabilise in the minimum energy configuration and dynamically minimise the eDM; this is exactly what we see in a CO$_2$ molecule. This configuration, corresponding to this dynamical angle, is called the \ti{axion} and is a classical solution to the strong CP problem. The strong CP problem is solved and we are finished.

\newpage

\section{The Quantum Solution To The Strong CP Problem}

Unfortunately, we have not got off that easy, for there is more to our world than classical physics. We must formulate and solve the strong CP problem at the quantum level.

\subsection{Chiral Symmetry}\label{Chiral Symmetry}

Consider an $SU(3)$ gauge theory with one flavour of massless quark. The quark is represented by the Dirac fermion field in the fundamental representation of the gauge group $SU(3)$: \begin{equation}\label{Dirac} \psi=\begin{pmatrix} \psi_L\\ \psi_R\end{pmatrix}\end{equation} where $\psi_L$ and $\psi_R$ are left- and right-handed Weyl fields, two-component spinors. The vector gauge boson is called the \ti{gluon} and is the quanta of the gauge field $A_\mu$.\\

The classical QCD Lagrangian for one flavour of massless quark reads: \begin{equation}\label{QCD Lagrangian} \Lie=i \bar{\psi} \slashed{D} \psi-\frac{1}{4} F^{\mu \nu,a} F_{\mu \nu}^{a}-\frac{g^{2} \theta}{32 \pi^{2}} \tilde{F}^{\mu \nu,a} F_{\mu \nu}^{a}\end{equation} where $F_{\mu \nu}^{a}=\partial_{\mu} A_{\nu}^{a}-\partial_{\nu} A_{\mu}^{a}+g f^{a b c} A_{\mu}^{b} A_{\nu}^{c}$ is the $SU(3)$ valued field strength tensor for the strong interaction, $\tilde{F}^{\mu \nu,a}=\frac{1}{2} \epsilon^{\mu \nu \rho \sigma} F^a_{\rho \sigma}$ is it's Hodge dual, and $\slashed{D}:=\gamma^\mu (\p_\mu-igA_\mu)$ is the covariant derivative contracted with the usual gamma matrices. We have also introduced the strong coupling constant $g$ and an arbitrary parameter $\theta$ which parameterises a CP violating term. In case the reader is more familiar with seeing the gauge fields inside a trace, we note the two notations can be easily moved between by expanding the gauge fields in terms of generator matrices: \begin{equation} \text{Tr}[\tilde{F}^{\mu\nu}F_{\mu\nu}]=\text{Tr}[\tilde{F}^{\mu\nu,a}T^aF_{\mu\nu}^bT^b]=\tilde{F}^{\mu\nu,a}F_{\mu\nu}^b\text{Tr}[T^aT^b]=\tilde{F}^{\mu\nu,a}F_{\mu\nu}^bT(N)\delta^{ab}=\frac{1}{2}\tilde{F}^{\mu\nu,a}F_{\mu\nu}^a\end{equation} where we have substituted the \ti{Index} $\operatorname{Tr}\left(T_{R}^{a} T_{R}^{b}\right)=T(R) \delta^{a b}$ for some representation $R$ and used that in the fundamental representation $T(N)=\frac{1}{2}$.\\

In addition to the SU(3) gauge symmetry, this classical Lagrangian has a \ti{global} symmetry \begin{equation}G=U(1)_{V} \times U(1)_{A}\end{equation} where $U(1)_V$ is a \ti{vector} $U(1)$ symmetry, and $U(1)_A$ is an \ti{axial} $U(1)$ symmetry.\\

The `vector' and `axial' distinction on these last two global symmetries might be less familiar to the reader, to help explain their origin, consider the Lagrangian (\ref{QCD Lagrangian}) in two-component form: \begin{equation}\label{2-component} \Lie=i \psi^\dag_L\bar{\sigma}_\mu D^\mu \psi_L+i \psi^\dag_R\sigma_\mu D^\mu \psi_R-\frac{1}{4} F^{\mu \nu,a} F_{\mu \nu}^{a}-\frac{g^{2} \theta}{32 \pi^{2}} \tilde{F}^{\mu \nu,a} F_{\mu \nu}^{a}\end{equation} where $\sigma:=\left(\mbb{1}_2, \sigma^{i}\right)$ and $\bar{\sigma}:=\left(\mbb{1}_2,-\sigma^{i}\right)$ with the Pauli spin matrices $\sigma^i$; these act as the 2-component equivalent of the gamma matrices. The Lagrangian can thus possess two types of global $U(1)$ symmetry: \begin{equation}\label{U(1) symmetries} \begin{aligned} U(1)_V:&\quad \psi_L\to e^{i\alpha}\psi_L,\quad &\psi_{R}\to e^{i\alpha}\psi_R\quad&\equiv\quad \psi\to e^{i\alpha}\psi\\
U(1)_A:&\quad \psi_L\to e^{-i\alpha}\psi_L,\quad &\psi_{R}\to e^{i\alpha}\psi_R\quad&\equiv\quad \psi\to e^{i\alpha\gamma^5}\psi\end{aligned}\end{equation} recalling the \ti{chirality operator} $\gamma_{5} \psi_{L}= -\psi_{L},\;\gamma_{5} \psi_{R}= \psi_{R}$. The former is called a \ti{vector} $U(1)$ symmetry because the associated Noether current $J_V^{\mu}(x) \equiv \bar{\psi} \gamma^{\mu} \psi(x)$ transforms as a vector. The latter is called an \ti{axial} $U(1)$ symmetry because the associated Noether current $J_A^{\mu}(x) \equiv \bar{\psi} \gamma^{\mu}\gamma^5 \psi(x)$ transforms as an axial vector (the spacial part is odd under parity). The $U(1)_A$ symmetry treats left- and right-handed Weyl fields differently and is thus known as a \ti{chiral symmetry}. Note the $U(1)_A$ transformation law for $\bar{\psi}$: \begin{equation}\label{anti xform}U(1)_A:\quad \bar{\psi}=\psi^\dag\gamma^0\to\left(e^{i\alpha\gamma^5}\psi\right)^\dag\gamma^0=\psi^\dag e^{-i\alpha(\gamma^5)^\dag}\gamma^0=\psi^\dag\gamma^0 e^{i\alpha\gamma^5}=\bar{\psi}e^{i\alpha\gamma^5}\end{equation} is the same as the $U(1)_A$ transformation law for $\psi$, recalling that $(\gamma^5)^\dag=\gamma^5$ and $\{\gamma^5,\gamma^\mu\}$=0. The remaining conjugate transformations are obvious (just flip the sign on the $i$).

\subsection{Anomalous Symmetries} \label{Anom}

We now consider how these symmetries behave upon adding a mass term for the quark and quantising the theory. We start by constructing a mass term for the quark, which must be real, Lorentz invariant, and gauge invariant. Lorentz and gauge invariance are only satisfied if we pair up left- and right-handed Weyl spinors, leading us to write down a mass term \begin{equation}\label{quark mass}\Lie_\text{Mass} \overset{?}{=} -m\psi^\dag_L\psi_R.\end{equation} However, the fermion masses and the components of the Weyl spinors are in general complex-valued, so to obtain a \ti{real}, Lorentz invariant, and gauge invariant mass term we must write (\ref{quark mass}) along with its complex conjugate, yielding a quark mass term \begin{equation}\label{quark mass term} \Lie_\text{Mass}=-m\psi^\dag_L\psi_R-m^*\psi^\dag_R\psi_L.\end{equation} Explicitly pulling out the arbitrary complex phase of the mass: \begin{equation} m=|m|e^{i\theta_q}\end{equation} allows us to write (\ref{quark mass term}) in terms of Dirac fields: \begin{equation}\label{quark mass term 2} \Lie_\text{mass}=-|m|\bar{\psi}e^{i\theta_q\gamma^5}\psi.\end{equation} We observe that applying a $U(1)_A$ transformation \begin{equation}\label{mass xform} U(1)_A:\quad-|m|\bar{\psi}e^{i\theta_q\gamma^5}\psi \to-|m|\bar{\psi}e^{i(\theta_q+2\alpha)\gamma^5}\psi \end{equation} does not leave the mass term invariant. We thus discover that $U(1)_A$ is only an \ti{approximate} symmetry of the classical theory which becomes an exact symmetry in the limit of massless quarks.\\

To quantise this theory, we make use of the path integral: \begin{equation}\label{path integral} Z=\int \mathcal{D}A\mathcal{D}\bar{\psi}\mathcal{D}\psi e^{i\int d^4x (i \bar{\psi} \slashed{D} \psi-\frac{1}{4} F^{\mu \nu,a} F_{\mu \nu}^{a}-\frac{g^{2} \theta}{32 \pi^{2}} \tilde{F}^{\mu \nu,a} F_{\mu \nu}^{a})}\end{equation} excluding source terms for brevity. From the Fujikawa method \cite{Fuj}, it can be shown that the functional measure of the path integral is not invariant under a $U(1)_A$ transformation, but rather transforms as \begin{equation}\label{measure}\mathcal{D}A\mathcal{D}\bar{\psi}\mathcal{D}\psi\to \mathcal{D}A\mathcal{D}\bar{\psi}\mathcal{D}\psi \exp \left[i \int \alpha\frac{g^2}{16 \pi^{2}} \tilde{F}^{\mu \nu,a} F_{\mu \nu}^{a} d x\right]\end{equation} resulting in an additional term in the Lagrangian: \begin{equation}\label{Fuj term} \Lie_\text{Fuj}=\alpha \frac{g^{2}}{16 \pi^{2}} \tilde{F}^{\mu \nu,a} F_{\mu \nu}^{a}\end{equation} and thus a $U(1)_A$ transformation does not leave the quantised theory invariant. When a given symmetry is a good symmetry of the classical theory but not a good symmetry of the quantum theory it is referred to as an \ti{anomalous symmetry}. Since $U(1)_A$ is an approximate symmetry of the classical theory but not a symmetry at all of the quantised theory, $U(1)_A$ is an \ti{anomalous symmetry}.\\

Considering these two effects together, the \ti{quantised} QCD Lagrangian for one flavour of \ti{massive} quark transforms under the $U(1)_A$ transformation as: \begin{equation}\label{QCD Lagrangian xform}\begin{aligned} \Lie_\text{QCD}&=i \bar{\psi} \slashed{D} \psi-|m|\bar{\psi}e^{i\theta_q\gamma^5}\psi-\frac{1}{4} F^{\mu \nu,a} F_{\mu \nu}^{a}-\frac{g^{2} \theta}{32 \pi^{2}} \tilde{F}^{\mu \nu,a} F_{\mu \nu}^{a}\\
&\to i \bar{\psi} \slashed{D} \psi-|m|\bar{\psi}e^{i(\theta_q+2\alpha)\gamma^5}\psi-\frac{1}{4} F^{\mu \nu,a} F_{\mu \nu}^{a}-(\theta-2\alpha)\frac{g^{2}}{32 \pi^{2}} \tilde{F}^{\mu \nu,a} F_{\mu \nu}^{a}\end{aligned}\end{equation} and is no longer invariant. $U(1)_A$ is an \ti{anomalous symmetry}; it is no longer a good symmetry of nature. Note that in all cases we still have invariance under a $U(1)_V$ transformation and thus it is a good symmetry of nature. We may be tempted to just throw $U(1)_A$ away, but that would be too hasty, there is something we can salvage from this apparent tragedy.

\subsection{Spurions}

We can promote the constants $\theta_q$ and $\theta$ to spacetime dependent, fictions, auxiliary fields; $\theta_q$ and $\theta$ are then called \ti{spurions}. We give them the following $U(1)_A$ transformations: \begin{equation} U(1)_A:\quad \theta_q\to\theta-2\alpha, \quad \theta\to\theta+2\alpha\end{equation} which cancels the $\Lie_\text{Fuj}$ term and restores the quark mass term to its original form in the $U(1)_A$ transformation of the quantised QCD Lagrangian for one flavour of massive quark (\ref{QCD Lagrangian xform}). We thus find that, when introducing spurions, the $U(1)_A$ transformation: \begin{equation}\label{1 flavour anom sym} \psi\to e^{i\alpha\gamma^5}\psi,\quad \theta_q\to\theta_q-2\alpha,\quad \theta\to\theta+2\alpha\end{equation} is a good symmetry of the quantised QCD Lagrangian for one flavour of massive quark.\\

The spurion fields are fictions, auxiliary fields and thus after all is said and done we must set the spurious fields equal to the constants $\theta_q$ and $\theta$ to get our physical theory. $U(1)_A$ itself will never be a good symmetry of nature but the above transformation (\ref{1 flavour anom sym}), which we will call a \ti{spurious symmetry}, is.\\

A final but important note: any physical quantities can only depend on parameters that are invariant under any good symmetry of nature, such as the above spurious symmetry (\ref{1 flavour anom sym}). Thus, any physical quantity cannot depend on $\theta$ or $\theta_q$ alone, but only on the combination $\theta+\theta_q$; clearly invariant under (\ref{1 flavour anom sym}).

\subsection{Two Flavour QCD}

Recall our objective, investigating the strong CP problem. Since the strong CP problem's simplest manifestations are in the mass of the pions and the eDM of the neutron, we need to investigate the quantum theory describing neutrons and pions. Since QCD is a theory describing quarks, a low-energy QCD effective field theory (EFT) will describe neutrons (\ti{nucleons} to be exact) which, as we will discover, interact via pions. We thus consider QCD with 2 light flavours of quarks. In the previous section we had a single quark represented by the Dirac field $\psi$. From now on we have \ti{two} quarks represented by the Dirac fields $u$ and $d$: \begin{equation} u=\begin{pmatrix}u_L\\u_R\end{pmatrix}\quad\quad d=\begin{pmatrix}d_L\\d_R\end{pmatrix}\end{equation} where $u_L,d_L$ and $u_R,d_R$ are left- and right-handed Weyl fields, two component spinors. It will be helpful to define the Weyl field doublets: \begin{equation}\label{Weyl} \psi_L=\begin{pmatrix}u_L\\d_L\end{pmatrix}\quad\quad \psi_R=\begin{pmatrix}u_R\\d_R\end{pmatrix}.\end{equation} The QCD Lagrangian for one flavour of massive quark (\ref{QCD Lagrangian xform}) is easily extended to the case of two flavours of massive quark and reads: \begin{equation}\label{2 flavour L} \Lie_\text{QCD}=i \bar{u} \slashed{D} u-m_u\bar{u}e^{i\theta_u\gamma^5}u+i \bar{d} \slashed{D} d-m_d\bar{d}e^{i\theta_d\gamma^5}d-\frac{1}{4} F^{\mu \nu,a} F_{\mu \nu}^{a}-\frac{g^{2} \theta}{32 \pi^{2}} \tilde{F}^{\mu \nu,a} F_{\mu \nu}^{a}\end{equation} where just as before, the up and down quark each have a complex mass with arbitrary phases $\theta_u$ and $\theta_d$ respectively. Considering the Lagrangian (\ref{2 flavour L}) in 2-component form, the Dirac kinetic terms read: \begin{equation} \Lie_\text{QCD}\supset i \bar{u} \slashed{D} u+i \bar{d} \slashed{D} d=i \psi^\dag_L\bar{\sigma}_\mu D^\mu \psi_L+i \psi^\dag_R\sigma_\mu D^\mu \psi_R\end{equation} making use of the Weyl field doublets (\ref{Weyl}). We thus observe a manifest $SU(2)_L\times SU(2)_R$ global 2-flavour symmetry for the left- and right-handed Weyl field doublets: \begin{equation}\label{SU(2) xform} \begin{aligned} SU(2)_L:& \quad\psi_L\to L\psi_L\\
SU(2)_R:& \quad\psi_R\to R\psi_R\end{aligned}\end{equation} where $L\in SU(2)_L$, $R\in SU(2)_R$, and thus $L^\dag L=\mbb{1}_2$, $R^\dag R=\mbb{1}_2$. However, the Dirac mass terms read: \begin{equation}\label{2-flav}\begin{aligned} \Lie_\text{QCD}\supset& -m_u\bar{u}e^{i\theta_u\gamma^5}u-m_d\bar{d}e^{i\theta_d\gamma^5}d\\
=&-m_uu^\dag_Lu_R-m_u^*u^\dag_Ru_L-m_dd^\dag_Ld_R-m_d^*d^\dag_Rd_L\end{aligned}\end{equation} using complex quark masses for ease of notation. This is clearly not invariant under an $SU(2)_L\times SU(2)_R$ transformation due to the quark masses. For the case of equal quark masses ($m_u=m_d=m$) (\ref{2-flav}) becomes: \begin{equation} -m\psi^\dag_L\psi_R-m^*\psi^\dag_R\psi_L\end{equation} which transforms under an $SU(2)_L\times SU(2)_R$ transformation as: \begin{equation} SU(2)_L\times SU(2)_R:\quad-m\psi^\dag_L\psi_R-m^*\psi^\dag_R\psi_L\to-m\psi^\dag_LL^\dag R\psi_R-m^*\psi^\dag_RR^\dag L\psi_L.\end{equation} This is only invariant for the case of $L\equiv R$, we call this a \ti{vector} $SU(2)_L\times SU(2)_R$ transformation. The case of $L\neq R$ is called an \ti{axial} $SU(2)_L\times SU(2)_R$ transformation which our theory is only invariant under for \ti{massless} quarks. To aid in our description we can equivalently write: $SU(2)_L\times SU(2)_R\equiv SU(2)_V\times SU(2)_A$ where $SU(2)_V$ and $SU(2)_A$ are vector and axial $SU(2)_L\times SU(2)_R$ transformations. The $SU(2)_V$ and $SU(2)_A$ symmetries are thus only \ti{approximate} symmetries of our theory which become exact symmetries in the equal quark mass and massless quark limits respectively; the former is known as \ti{isospin symmetry}. A final note: the differing electromagnetic charge of the quarks also contributes to the approximate nature of these symmetries; QED breaks $SU(2)_L\times SU(2)_R$ symmetry. Since we only consider QCD this will not play an important role, but is mentioned for completeness.\\

Our theory of QCD for two light flavours of quarks is thus \ti{approximately} invariant under the global symmetries $SU(2)_V\times SU(2)_A$ and \ti{exactly} invariant under the global symmetry $U(1)_V$ and the \ti{two} transformations: \begin{equation}\label{anom sym}\begin{aligned} u&\to e^{i\alpha\gamma^5}u,\quad \theta_u&\to\theta_u-2\alpha,\quad \theta&\to\theta+2\alpha\\
d&\to e^{i\alpha\gamma^5}d,\quad \theta_d&\to\theta_d-2\alpha,\quad \theta&\to\theta+2\alpha\end{aligned}\end{equation} which are just two copies of the spurious symmetry we found in the previous section (\ref{1 flavour anom sym}), one for each quark flavour. $\theta_u,\theta_d$, and $\theta$ are spurions.

\subsection{Chiral Symmetry Breaking} \label{Chiral Symmetry Breaking}

It is an experimental observation that the quark condensate has a non-zero vacuum expectation value (vev): \begin{equation} \label{vev} \langle \psi_R\psi_L^\dag\rangle=-v^3\mbb{1}_2\end{equation} and thus the approximate $SU(2)_{L} \times SU(2)_{R}$ symmetry is \ti{spontaneously broken}, as a preferred direction of the quark doublets is chosen while in the ground state. This process is called \ti{spontaneous symmetry breaking} (SSB). A field which is not turned off in the vacuum is totally unphysical and indeed will not correspond to physical states when the system undergoes SSB. We thus perform a \ti{field expansion} of the quark condensate around its non-zero vev: \begin{equation} \label{vev 2} \psi_R\psi_L^\dag=-v^3U(x)\end{equation} where $U(x)$ is a spacetime dependent unitary matrix, defined as \begin{equation}\label{U matrix} U(x):=e^{\frac{i\pi^a(x)\sigma^{a}}{f_\pi}}e^{i\theta}\end{equation} where $\pi^a$ are a set of fields with zero vev ($\langle\pi^a\rangle=0$), $\sigma^a$ are a corresponding set of matrices, $f_\pi$ is a constant with dimensions of mass, and $\theta$ is our spurious field which we include so that the LHS and RHS of (\ref{vev 2}) transform identically under the spurious symmetry (\ref{anom sym}), as they must: \begin{equation} \begin{aligned} \psi_R\psi_L^\dag \to& \psi_R\psi_L^\dag e^{2i\alpha} \\
-v^3U(x)\to&-v^3U(x)e^{2i\alpha}\end{aligned}\end{equation} using the $U(1)_A$ transformations of the left- and right-handed Weyl fields (\ref{U(1) symmetries}).\\

Since the fields $\pi^a$ have zero vev they will describe the physical states of our system after SSB, the particles we would detect. To determine the number and nature of these particles we need to investigate the details of the symmetry breaking.\\

Applying an $SU(2)_L\times SU(2)_R$ transformation to the vev $(\ref{vev})$ yields: \begin{equation}\label{vev xform} SU(2)_L\times SU(2)_R:\quad -v^3\mbb{1}_2\to -v^3(RL^\dag)\end{equation} using the transformations in (\ref{SU(2) xform}). Observe that, for the case of $L= R$, the vev is unchanged under an $SU(2)_L\times SU(2)_R$ transformation, but if $L\neq R$, the vev is changed. Thus, $SU(2)_A$ is \ti{spontaneously broken} but the $SU(2)_V$ symmetry remains.\\

Applying $U(1)_V$ and $U(1)_A$ transformations to the vev $(\ref{vev})$ yields: \begin{equation}\label{vev xform U(1)}\begin{aligned} U(1)_V:&\quad -v^3\mbb{1}_2\to -v^3e^{i\alpha}\mbb{1}_2e^{-i\alpha}=-v^3\\
U(1)_A:&\quad -v^3\mbb{1}_2\to -v^3e^{i\alpha}\mbb{1}_2e^{i\alpha}=-v^3e^{2i\alpha}\end{aligned}\end{equation} using the transformations for left- and right-handed Weyl fields (\ref{U(1) symmetries}). We observe the vev to be unchanged under a $U(1)_V$ transformation but changed under a $U(1)_A$ transformation. Thus, the anomalous $U(1)_A$ symmetry is \ti{spontaneously broken} but the $U(1)_V$ symmetry remains.\\

Goldstones therom \cite{Goldstone} states:\\
\\
\ti{Whenever a continuous symmetry of the Lagrangian is spontaneously broken, massless `Goldstone bosons' emerge, with one present for each broken generator of the symmetry.}\\
\\
However, the spontaneously broken symmetry $SU(2)_A$ was only ever an approximate symmetry of our theory. The result of which is that any corresponding Goldstone bosons arising from SSB of this symmetry are not exactly massless, as Goldstone bosons for perfect symmetries are. We call these almost-massless Goldstone bosons \ti{pseudo-Goldstone bosons}. Although $U(1)_A$ is an approximate symmetry of the classical theory, recall it is not a symmetry at all of the quantised theory; it is anomalous. Since there is no symmetry at all there will be no corresponding Goldstone boson after SSB, since there was never a symmetry to break in the first place.\\

We must now identify the broken generators to find our pseudo-Goldstone bosons. Recalling that $SU(N)$ has $N^2-1$ generators, the breaking of $SU(2)_A$ will have $2^2-1=3$ broken generators and thus has 3 corresponding pseudo-Goldstone bosons. We thus determine that $a=1,2,3$ where $\sigma^a$ are the Pauli spin matrices and $\pi^a$ are the pseudo-Goldstone bosons associated with the breaking of $SU(2)_A$. These are related to the pions: \begin{equation}\label{explicit pions}\pi^0=\pi^3,\quad \pi^\pm=\frac{1}{\sqrt{2}}(\pi^1\pm i\pi^2).\end{equation} The value of $f_\pi$ can be determined from the rate of decay of a $\pi^+$ via the weak interaction and is thus called the \ti{pion decay constant} with value \cite{PDG} \begin{equation}\label{fpi} f_\pi \approx130\text{ MeV}. \end{equation}\\

We've extracted a lot of information about the symmetries of our theory so before proceeding we present a summary of the nature of the symmetries encountered:

\begin{itemize}
\item $SU(2)_V$: Approximate symmetry (isospin).
\item $SU(2)_A$: Approximate symmetry, spontaneously broken.
\item $U(1)_V$: $\;\;$Good symmetry.
\item $U(1)_A$:  $\;\;$Approximate symmetry, spontaneously broken, anomalous.
\end{itemize}
Recalling also that the spurious symmetries (\ref{anom sym}) are good symmetries of our theory.

\subsection{Low-Energy QCD} \label{Low-Energy QCD}

$U(x)$ will act as the effective field of our low-energy QCD EFT. To build the effective Lagrangian, we consider all possible terms invariant under the remaining symmetries of our theory: $SU(2)_V\times U(1)_V$ and the spurious symmetries (\ref{anom sym}). Note that since $U(1)_V$ acts trivially on $U(x)$ we need not explicitly consider it.

\subsubsection{Pion Kinetic Term} \label{pions}

We first consider a kinetic term for the pions. A general $SU(2)_L\times SU(2)_R$ transformation of $U(x)$ yields: \begin{equation}\label{U(x) xform} SU(2)_L\times SU(2)_R: \quad U(x)\to RU(x)L^\dag\end{equation} and thus an $SU(2)_V$ transformation will act just as (\ref{U(x) xform}) but limited to the case of $L=R$.\\

Using the cyclicality of the trace and $R^\dag L=\mbb{1}_2$, the simplest term we can think of that is invariant under an $SU(2)_V$ transformation and the spurious symmetries (\ref{anom sym}) is \begin{equation} \Lie\supset \text{Tr}\left[U^\dag U\right].\end{equation} However, since $U^\dag U=\mbb{1}_2$, this term only contributes as an additional constant which can always be removed from the Lagrangian.\\

The next term we can think of will involve derivatives and reads: \begin{equation}\label{U Lagrangian} \mathcal{L}\supset c\operatorname{Tr}\left[ \partial^{\mu} U^{\dagger} \partial_{\mu} U\right]\end{equation} where $c$ is an arbitrary coefficient. Since the $SU(2)_V$ transformation is a \ti{global} transformation, $L$ and $R$ will commute with the partial derivatives $\p^\mu$ and thus (\ref{U Lagrangian}) is indeed invariant. To get something we can work with on the level of Feynman rules we expand $U$ (\ref{U matrix}) in powers of $f^{-1}_\pi$: \begin{equation}\label{U expansion}\begin{aligned}U&= \left(\mbb{1}_2+\frac{i\pi^a\sigma^a}{f_\pi}-\frac{\pi^a\sigma^a\pi^b\sigma^b}{2f^2_\pi}-\frac{i\pi^a\sigma^a\pi^b\sigma^b\pi^c\sigma^c}{6f^3_\pi}+\dots\right)e^{i\theta}\\
&=\left(\mbb{1}_2+\frac{i\pi^a\sigma^a}{f_\pi}-\frac{\pi^a\pi^a\mbb{1}_2}{2f^2_\pi}-\frac{i\pi^a\sigma^a\pi^b\pi^b}{6f^3_\pi}+\dots\right)e^{i\theta}\end{aligned}\end{equation} where we have used: \begin{equation}\label{pauli} \pi^a\sigma^a\pi^b\sigma^b=\frac{1}{2}\pi^a\pi^b\{\sigma^a,\sigma^b\}=\frac{1}{2}\pi^a\pi^b2\delta^{ab}=\pi^a\pi^a\mbb{1}_2\end{equation} making use of the pauli spin matrices anti-commutation relation $\{\sigma^a,\sigma^b\}=2\delta^{ab}$. Substituting the above expansion (\ref{U expansion}) into (\ref{U Lagrangian}) and working to leading order in $f_\pi^{-1}$: \begin{equation}\begin{aligned}\Lie&\supset c\text{Tr}\left[\p^\mu\left(\mbb{1}_2+\frac{i\pi^a\sigma^a}{f_\pi}-\frac{\pi^a\pi^a\mbb{1}_2}{2f^2_\pi}-\frac{i\pi^a\sigma^a\pi^b\pi^b}{6f^3_\pi}\right)^\dag\p_\mu\left(\mbb{1}_2+\frac{i\pi^c\sigma^c}{f_\pi}-\frac{\pi^c\pi^c\mbb{1}_2}{2f^2_\pi}-\frac{i\pi^c\sigma^c\pi^d\pi^d}{6f^3_\pi}\right)\right]\\
&= c\text{Tr}\Bigg[\left(-\frac{i\p^\mu\pi^a\sigma^a}{f_\pi}-\frac{\p^\mu\pi^b\pi^b\mbb{1}_2}{f^2_\pi}+i\frac{\p^\mu\pi^a\sigma^a\pi^b\pi^b+2\pi^a\sigma^a\p^\mu\pi^b\pi^b}{6f^3_\pi}\right)\\
&\quad\quad\;\times\left(\frac{i\p_\mu\pi^c\sigma^c}{f_\pi}-\frac{\p_\mu\pi^c\pi^c\mbb{1}_2}{f^2_\pi}-i\frac{\p_\mu\pi^c\sigma^c\pi^d\pi^d+2\pi^c\sigma^c\p_\mu\pi^d\pi^d}{6f^3_\pi}\right)\Bigg]\\
&=c\text{Tr}\left[\frac{1}{f_\pi^2}\p^\mu\pi^a\p_\mu\pi^b\sigma^a\sigma^b +\frac{1}{f_\pi^4}\p_\mu\pi^b\pi^b\p_\mu\pi^c\pi^c\mbb{1}_2-\frac{1}{3f^4_\pi}(\p^\mu\pi^a\sigma^a)(\p_\mu\pi^c\sigma^c\pi^d\pi^d+2\pi^c\sigma^c\p_\mu\pi^d\pi^d) \right]\\
&=c\text{Tr}\left[\frac{1}{f_\pi^2}\p^\mu\pi^a\p_\mu\pi^a\mbb{1}_2 +\frac{1}{f_\pi^4}\p_\mu\pi^b\pi^b\p_\mu\pi^c\pi^c\mbb{1}_2-\frac{1}{3f^4_\pi}(\p^\mu\pi^a\p_\mu\pi^a\pi^d\pi^d+2\p^\mu\pi^a\pi^a\p_\mu\pi^d\pi^d)\mbb{1}_2 \right]\\
&=c\text{Tr}\left[\frac{1}{f_\pi^2}\p^\mu\pi^a\p_\mu\pi^a\mbb{1}_2-\frac{1}{3f^4_\pi}(\p^\mu\pi^a\p_\mu\pi^a\pi^d\pi^d-\p^\mu\pi^a\pi^a\p_\mu\pi^d\pi^d)\mbb{1}_2 \right]\\
&=\frac{4c}{f^2_\pi}\left(\frac{1}{2}\p^\mu\pi^a\p_\mu\pi^a-\frac{1}{6f^2_\pi}(\p^\mu\pi^a\p_\mu\pi^a\pi^d\pi^d-\p^\mu\pi^a\pi^a\p_\mu\pi^d\pi^d) \right).\\
\end{aligned}\end{equation} We take $c=f^2_\pi/4$ such that the pion kinetic term is in the standard form for a real scalar field. Thus, to leading order in $f^{-1}_\pi$, the pion kinetic term and its naturally arising quartic interactions read: \begin{equation}\label{U Lagrangian 2} \Lie\supset \frac{1}{2}\p^\mu\pi^a\p_\mu\pi^a-\frac{1}{6f^2_\pi}(\p^\mu\pi^a\p_\mu\pi^a\pi^b\pi^b-\p^\mu\pi^a\pi^a\p_\mu\pi^b\pi^b).\end{equation} Although there are terms with more derivatives that we could include, these terms will be of $\Od(f^{-3}_\pi)$ and thus (\ref{U Lagrangian 2}) gives all possible leading order terms.\\

The ultraviolet (UV) cut-off for our EFT is $\Lambda\sim 4\pi f_\pi$ which makes $f^{-1}_\pi$ seem like a bad expansion parameter. However, it turns out that both tree and 1-loop diagrams yield approximately equal contributions when taking $(4\pi f_\pi)^{-1}$ rather than $f_\pi^{-1}$; the former is a good expansion parameter at the energy scales of our EFT and thus we only need to consider leading order terms.

\subsubsection{Pion Mass Term}

The quark mass terms in our 2 flavour QCD Lagrangian (\ref{2 flavour L}) can be written in terms of the quark doublet $\psi$ by the use of a mass matrix $\tilde{M}$: \begin{equation}\label{EFT mass term} \begin{aligned}\Lie&\supset -m_u\bar{u}e^{i\theta_u\gamma^5}u-m_d\bar{d}e^{i\theta_d\gamma^5}d\\
&=-m_ue^{i\theta_u}u^\dag_Lu_R - m_ue^{-i\theta_u}u^\dag_Ru_L-m_de^{i\theta_d}d^\dag_Ld_R - m_de^{-i\theta_d}d^\dag_Rd_L\\
&=-\psi^\dag_L\tilde{M}\psi_R - \psi^\dag_R\tilde{M}^\dag\psi_L\end{aligned}\end{equation} where the mass matrix $\tilde{M}$ is given by: \begin{equation} \tilde{M}=\left(\begin{array}{cc}m_{u}e^{i\theta_u} & 0 \\0 & m_{d}e^{i\theta_d}\end{array}\right). \end{equation} We are free to bring this mass matrix into the form \begin{equation}\label{mass matrix} SU(2)_V: \tilde{M}\to M=\left(\begin{array}{cc}m_{u} & 0 \\0 & m_{d}\end{array}\right)e^{i(\theta_u+\theta_d)}\end{equation} with just one overall phase via an $SU(2)_V$ transformation; which our theory is indeed invariant under.\\

Starting with (\ref{EFT mass term}) and using the cyclicality of the trace, we can replace the quark condensate $\psi_R\psi_L^\dag$ with its field expansion (\ref{vev 2}) and we find: \begin{equation}\label{EFT mass term 2}\begin{aligned} \Lie&\supset -\psi^\dag_LM\psi_R - \psi^\dag_RM^\dag\psi_L\\
&=-\text{Tr}\left[M\psi_R\psi^\dag_L+M^\dag\psi_L\psi^\dag_R\right]\\
&= v^3\text{Tr}\left[M U+M^\dag U^\dag\right]\\
&=-V(U)-V(U)^*\\
&=-2\text{Re}\left[V(U)\right]\end{aligned}\end{equation}
defining the potential \begin{equation}\label{theta potential} V(U):=-v^{3} \operatorname{Tr} \left[M U\right].\end{equation} The Mass of the pions will thus be given by: \begin{equation}\label{masses} m_{\pi^a}^2=2\text{Re}\left[\frac{\p^2 V(\pi^a)}{\p {\pi^a}^2}\mid_{\pi^a=\langle\pi^a\rangle}\right]\end{equation} where $\langle\pi^a\rangle$ are the vevs of the pions; this is simply the value of the pion fields at the minimum of the potential (\ref{EFT mass term 2}) and can thus be determined by solving \begin{equation} \label{vev equation} 2\text{Re}\left[\frac{\p V(\pi^a)}{\p {\pi^a}}\mid_{\pi^a=\langle\pi^a\rangle}\right]=0.\end{equation}\\

Substituting the explicit form of the spacetime dependent vev (\ref{U matrix}) into the potential (\ref{theta potential}) and expanding to leading order in $f^{-1}_\pi$: \begin{equation}\label{pi potential}
\begin{aligned} V(\pi^a)&=-v^3\text{Tr}\left[M\left(1+\frac{i\pi^a\sigma^a}{f_\pi}-\frac{\pi^a\sigma^a\pi^b\sigma^b}{2f^2_\pi}\right)e^{i\theta}\right]\\
&=-v^3\text{Tr}\left[M\left(1+\frac{i\pi^a\sigma^a}{f_\pi}-\frac{\pi^a\pi^a\mbb{1}_2}{2f^2_\pi}\right)e^{i\theta}\right]\end{aligned}\end{equation} making use of (\ref{pauli}). By substituting (\ref{pi potential}) into (\ref{vev equation}) we thus need to solve \begin{equation} 2\text{Re}\left[\frac{\p V(\pi^a)}{\p {\pi^a}}\mid_{\pi^a=\langle\pi^a\rangle}\right]=-2v^3\text{Re}\left[\text{Tr}\left[M\left(\frac{i\sigma^a}{f_\pi}-\frac{\langle\pi^a\rangle}{f^2_\pi}\right)e^{i\theta}\right]\right]=0. \end{equation} Observing that \begin{equation} \text{Tr}\left[M\sigma^a\right]=0\quad\quad \text{for }a=1,2\end{equation} we thus find \begin{equation}\label{pmvev}\langle\pi^1\rangle=\langle\pi^2\rangle=0\implies\langle\pi^\pm\rangle=0.\end{equation} However, for $a=3$: \begin{equation}\begin{aligned}
2\text{Re}\left[\frac{\p V(\pi^3)}{\p {\pi^3}}\mid_{\pi^3=\langle\pi^3\rangle}\right]&=-2v^3\text{Re}\left[\text{Tr}\left[M\left(\frac{i\sigma^3}{f_\pi}-\frac{\langle\pi^3\rangle}{f^2_\pi}\right)e^{i\theta}\right]\right]\\
&=-2v^3\text{Re}\left[\text{Tr}\left[\left(\frac{i}{f_\pi}\begin{pmatrix}m_u&0\\0&-m_d\end{pmatrix}-\frac{\langle\pi^3\rangle}{f^2_\pi}\begin{pmatrix}m_u&0\\0&m_d\end{pmatrix}\right)e^{i(\theta+\theta_u+\theta_d)}\right]\right]\\
&=-2v^3\text{Re}\left[\left(\frac{i}{f_\pi}(m_u-m_d)-\frac{\langle\pi^3\rangle}{f^2_\pi}(m_u+m_d)\right)e^{i\bar{\theta}}\right]=0\\
\end{aligned}\end{equation} where we have defined \begin{equation}\label{thetabar}\bar{\theta}:=\theta+\theta_u+\theta_d.\end{equation}\\

We note the parameter $\bar{\theta}$ is invariant under the spurious symmetries (\ref{anom sym}) and is thus a candidate for parameterising a physical quantity. Spoiler alert, $\bar{\theta}$ turns out to be the quantum analogue of the classical angle seen in figure \ref{Neutron}, which we know experimentally to be very small; we thus expand the exponential $e^{i\bar{\theta}}$ to leading order in $\bar{\theta}$. From this we find: \begin{equation} 2\text{Re}\left[\frac{\p V(\pi^3)}{\p {\pi^3}}\mid_{\pi^3=\langle\pi^3\rangle}\right]=-2v^3\text{Re}\left[\left(\frac{i}{f_\pi}(m_u-m_d)-\frac{\langle\pi^3\rangle}{f^2_\pi}(m_u+m_d)\right)\left(1+i\bar{\theta}\right)\right]=0\end{equation} and from the real part we finally arrive at: \begin{equation}\label{0vev} \langle\pi^3\rangle=\langle\pi^0\rangle=\bar{\theta} f_\pi\frac{m_d-m_u}{m_u+m_d}.\end{equation}\\

The above vev comprises our first cause for concern. (Pseudo-) Goldstone bosons should have zero vev, yet we find a non-zero vev of the $\pi^3$ pseudo-Goldstone boson. The vev is parametrised by the arbitrary parameter $\bar{\theta}$ and thus our only way out would be to set this to zero. This seems like a serious problem, requiring a `fine tuning' of our theory. Regardless, we should continue with our calculation of the pion masses.\\

Armed with these vevs, we can determine the pion masses to leading order in $f^{-1}_\pi$ and $\bar{\theta}$. Expanding (\ref{theta potential}) to leading order in $f_\pi^{-1}$: \begin{equation}
\begin{aligned} V(\pi^a)&=-v^3\text{Tr}\left[M\left(1+\frac{i\pi^a\sigma^a}{f_\pi}-\frac{\pi^a\sigma^a\pi^b\sigma^b}{2f^2_\pi} -\frac{i\pi^a\sigma^a\pi^b\sigma^b\pi^c\sigma^c}{6f^3_\pi}\right)e^{i\theta}\right]\\
&=-v^3\text{Tr}\left[M\left(1+\frac{i\pi^a\sigma^a}{f_\pi}-\frac{\pi^a\pi^a}{2f^2_\pi} -\frac{i\pi^a\pi^a\pi^c\sigma^c}{6f^3_\pi}\right)e^{i\theta}\right]\end{aligned}\end{equation} and substituting into (\ref{masses}) yields a mass: \begin{equation}\begin{aligned} m_{\pi^a}^2&= -2v^3\text{Re}\left[\text{Tr}\left[M\left(-\frac{1}{f^2_\pi} -\frac{i\langle\pi^a\rangle\sigma^a}{f^3_\pi}\right)e^{i\theta}\right]\right]\\
&=v^3\text{Re}\left[\frac{1}{f^2_\pi}(m_u+m_d)e^{i\bar{\theta}} +v^3\frac{i\langle\pi^a\rangle}{f^3_\pi}\text{Tr}\left[M\sigma^a\right]e^{i\theta}\right].\end{aligned}
\end{equation} Substituting the vevs (\ref{pmvev}), (\ref{0vev}), and expanding to leading order in $\bar{\theta}$ yields: \begin{equation}
\begin{aligned}m_{\pi^\pm}^2&=2v^3\text{Re}\left[\frac{1}{f^2_\pi}(m_u+m_d)(1+i\bar{\theta})\right]\\
&=\frac{2v^3}{f^2_\pi}(m_u+m_d)\end{aligned}\end{equation} and \begin{equation}\begin{aligned} m_{\pi^0}^2&=2v^3\text{Re}\left[\frac{1}{f^2_\pi}(m_u+m_d)(1+i\bar{\theta}) -v^3\frac{i\bar{\theta}}{f^2_\pi}\frac{(m_u-m_d)^2}{m_u+m_d}(1+i\bar{\theta})\right]\\
&=2\frac{v^3}{f_\pi}(m_u+m_d)\left[1+\bar{\theta}^2\frac{(m_u-m_d)^2}{(m_u+m_d)^2}\right].\end{aligned}
\end{equation}\\

We have found the $\pi^0$ mass to be dependent on $\bar{\theta}$. Since $\bar{\theta}$ is invariant under the spurious symmetries (\ref{anom sym}) it is totally okay for it to parameterise a physical quantity such as the pion mass, however, there is a crippling problem. Turning to our experimentalist friends it is an observed fact that $m_{\pi^\pm}\approx m_{\pi^0}$, where the lack of equality is due to loop effects in QED. Our results allow for the mass of the $\pi^0$ to differ from the mass of the charged pions for a non-zero value of $\bar{\theta}$, but there is no reason for it to be small. We thus observe the first manifestation of the strong CP problem at the quantum level, and it really is a problem.\\

We shall add the arrow that is this triumph to our quiver and proceed with our pursuit of a full low-energy QCD EFT, so that we may add another with the eDM of the neutron.

\subsubsection{Nucleon Terms}

We next look at terms involving nucleons. We define the nucleon field as the doublet \begin{equation}\label{nucleon doublet} N=\begin{pmatrix}p\\n\end{pmatrix}\end{equation} where $p=uud$ is the proton and $n=udd$ is the neutron; all the objects present are Dirac fermions. Under a general $SU(2)_L\times SU(2)_R$ transformation, the nucleon field transforms as: \begin{equation} \begin{array}{l}
P_{L} N \rightarrow L P_{L} N \\
P_{R} N \rightarrow R P_{R} N
\end{array}\end{equation} where $P_{L,R}:=\frac{1}{2}(\mbb{1}_2\mp\gamma^5)$ is a projector picking out either the left- or right-handed part of $N$. Using $\{\gamma^\mu,\gamma^5\}=0$, $(\gamma^5)^\dag=\gamma^5$, and $(\gamma^5)^2=\mbb{1}_2$, it is easy to verify the following properties of $P_{L,R}$: \begin{equation}\label{Projector}\begin{aligned} &P_{L,R}\gamma^0\equiv\gamma^0P_{R,L},\\
&P_{L,R}^\dag\equiv P_{L,R},\\
&(P_{L,R})^2\equiv P_{L,R},\\
&P_{L,R}P_{R,L}\equiv0. \end{aligned}\end{equation}\\

An $SU(2)_V$ transformation will leave the standard Dirac kinetic term $i \bar{N} \slashed{\partial} N$ invariant, but not the standard Dirac mass term $m_N\overline{N}N$. However, an invariant mass term can be constructed by including appropriate factors of $U$ and $U^\dag$ (\ref{U matrix}) and reads: \begin{equation} \mathcal{L}\supset-m_{N} \overline{N}\left(U^{\dagger} P_{\mathrm{L}}+U P_{\mathrm{R}}\right) N.\end{equation}\\

There is one other parity, time-reversal, $SU(2)_V$, and (\ref{anom sym}) invariant term with only one derivative (recall that we are working to leading order in $f_\pi^{-1}$ thus only consider terms with one derivative). Including this, we have all the relevant terms involving nucleons: \begin{equation}\label{nucleon L}
\mathcal{L}\supset i \bar{N} \slashed{\partial} N-m_{N} \bar{N}\left(U^{\dagger} P_{L}+U P_{R}\right) N -\frac{1}{2}\left(\lambda-1\right) i \bar{N} \gamma^{\mu}\left(U \partial_{\mu} U^{\dagger} P_{{L}}+U^{\dagger}\partial_{\mu} U P_{{R}}\right) N \end{equation} where \begin{equation}\label{lambda} \lambda:=-\frac{g_A}{g_v}=1.27 \end{equation} is the \ti{ratio of axial-vector to vector coupling} with its value determined from the decay rate of the neutron via the weak interaction \cite{Mendenhall}.\\

We can tidy this up a bit by performing a field redefinition of the nucleon field: \begin{equation}\label{Nucleon field redefinition} \mathcal{N} :=\left(u^{\dagger} P_{\mathrm{L}}+u P_{\mathrm{R}}\right) N\end{equation} where $u^2:=U$; equivalently, using $u^\dag u=\mbb{1}_2$, we have: \begin{equation}\label{nucleon}N=\left(u P_{\mathrm{L}}+u^{\dagger} P_{\mathrm{R}}\right) \mathcal{N}.\end{equation} Making use of the identities: \begin{equation}\begin{aligned}&\partial_{\mu} U\equiv\left(\partial_{\mu} u\right) u+ u\left(\partial_{\mu} u\right),\\&\left(\partial_{\mu} u^{\dagger}\right) u\equiv-u^{\dagger}\left(\partial_{\mu} u\right),\end{aligned}\end{equation} and the relations in (\ref{Projector}), we can substitute (\ref{nucleon}) into (\ref{nucleon L}) and ultimately obtain: \begin{equation} \mathcal{L}\supset i \overline{\mathcal{N}} \slashed{\partial} \mathcal{N}-m_{N} \overline{\mathcal{N}} \mathcal{N}+\overline{\mathcal{N}} \slashed{a} \mathcal{N}-\lambda \overline{\mathcal{N}} \slashed{b} \gamma_{5} \mathcal{N}\end{equation} where we define the hermitian vector fields: \begin{equation}\begin{aligned}
a_{\mu} & := \frac{1}{2} i\left[u^{\dagger}\left(\partial_{\mu} u\right)+u\left(\partial_{\mu} u^{\dagger}\right)\right], \\
b_{\mu} & := \frac{1}{2} i\left[u^{\dagger}\left(\partial_{\mu} u\right)-u\left(\partial_{\mu} u^{\dagger}\right)\right].
\end{aligned}\end{equation} One again, expanding $u$ and $u^\dag$ to leading order in $f_\pi^{-1}$ yields: \begin{equation}\mathcal{L}\supset i \overline{\mathcal{N}} \slashed{\partial} \mathcal{N}-m_{N} \overline{\mathcal{N}} \mathcal{N}+\frac{\lambda}{2 f_{\pi}} \partial_{\mu} \pi^{a} \overline{\mathcal{N}} \sigma^{a} \gamma^{\mu} \gamma_{5} \mathcal{N}.\end{equation}\\

As it turns out, the pion-nucleon interaction term is one of the two terms used when calculating the neutron eDM; it will be helpful to simplify it a bit. Granting it centre stage: \begin{equation}\label{piNN}\begin{aligned}
\Lie_{\pi\bar{\mathcal{N}}\mathcal{N}}&=\frac{\lambda}{2 f_{\pi}} \partial_{\mu} \pi^{a} \overline{\mathcal{N}} \sigma^{a} \gamma^{\mu} \gamma_{5} \mathcal{N}\\
&=-\frac{\lambda}{2 f_{\pi}} \pi^{a} \partial_{\mu}\left(\overline{\mathcal{N}} \sigma^{a} \gamma^{\mu} \gamma_{5} \mathcal{N}\right)\\
&=-\frac{\lambda}{2 f_{\pi}} \pi^{a}\left(\slashed{\p}\overline{\mathcal{N}} \sigma^{a} \gamma_{5} \mathcal{N}-\overline{\mathcal{N}} \sigma^{a} \gamma_{5} \slashed{\p}\mathcal{N}\right)\\
&=-\frac{\lambda}{2 f_{\pi}} \pi^{a}\left(im_N\overline{\mathcal{N}} \sigma^{a} \gamma_{5} \mathcal{N}+\overline{\mathcal{N}} \sigma^{a} \gamma_{5} m_N\mathcal{N}\right)\\
&=-ig_{\pi\bar{N}N}\pi^a\overline{\mathcal{N}} \sigma^{a} \gamma_{5} \mathcal{N}
\end{aligned}\end{equation} where we integrate by parts in the second line, anti-commute $\slashed{\p}$ and $\gamma^5$ in the third line, substitute the Dirac equations $(i\slashed{p}-m)\mathcal{N}=0$ and $(i\slashed{p}+m)\overline{\mathcal{N}}=0$ in the forth line, and define the pion-nucleon coupling
constant: \begin{equation} g_{\pi\bar{N}N}:=\frac{\lambda m_N}{f_\pi}\end{equation} in the last line. Note the use of the Dirac equations is only applicable for on-shell nucleons; an approximation we will later see to be justified. Substituting $f_\pi =130\text{ MeV}$, $\lambda=1.27$, and $m_n=939.6$ GeV the pion-nucleon coupling constant takes the value: \begin{equation}\label{pnn} g_{\pi\bar{N}N}=9.2.\end{equation}

\subsubsection{Effective Lagrangian}

The last type of term we could write down will involve all of the above, $N$'s, $U$'s, and $M$'s. There are exactly 3 which are parity, time-reversal, $SU(2)_V$, and (\ref{anom sym}) invariant with no derivatives that are bilinear in the nucleon field $N$ and have one factor of the quark mass matrix $M$. Writing them out with arbitrary coefficients: \begin{equation}\begin{aligned}\Lie\supset &-c_{1} \overline{N}\left(M P_{\mathrm{L}}+M^{\dagger} P_{\mathrm{R}}\right) N\\
&-c_{2} \overline{N}\left(U^{\dagger} M^{\dagger} U^{\dagger} P_{\mathrm{L}}+U M U P_{\mathrm{R}}\right) N \\
&-c_{3} \operatorname{Tr}\left(M U+M^{\dagger} U^{\dagger}\right) \bar{N}\left(U^{\dagger} P_{\mathrm{L}}+U P_{\mathrm{R}}\right) N \\
&-c_{4} \operatorname{Tr}\left(M U-M^{\dagger} U^{\dagger}\right) \bar{N}\left(U^{\dagger} P_{\mathrm{L}}-U P_{\mathrm{R}}\right) N.
\end{aligned}\end{equation} Making the same field redefinition (\ref{nucleon}) takes us to: \begin{equation}\begin{aligned}\label{additional terms} \Lie\supset &-\frac{1}{2} c_{+} \overline{\mathcal{N}}\left[uM u+u^{\dagger}M^{\dagger} u^{\dagger}\right] \mathcal{N} \\
&+\frac{1}{2} c_{-} \overline{\mathcal{N}}\left[uM u-u^{\dagger}M^{\dagger} u^{\dagger}\right] \gamma_{5} \mathcal{N} \\
&-c_{3} \operatorname{Tr}\left[M {U}+M^{\dagger} {U}^{\dagger}\right] \overline{\mathcal{N}} \mathcal{N} \\
&+c_{4} \operatorname{Tr}\left[M {U}-M^{\dagger} {U}^{\dagger}\right] \overline{\mathcal{N}} \gamma_{5} \mathcal{N}
\end{aligned}\end{equation} where $c_{\pm}:=c_1\pm c_2$. As always, we expand the above to leading order in $f^{-1}_\pi$ and $\bar{\theta}$ which ultimately yields: \begin{equation} \mathcal{L}\supset-i \theta \mu\left(c_{-}+4 c_{4}\right) \overline{\mathcal{N}} \gamma_{5} \mathcal{N}-\frac{\bar{\theta} c_{+} \mu}{f_{\pi}} \pi^{a} \overline{\mathcal{N}} \sigma^{a} \mathcal{N}\end{equation} defining the reduced mass of the quarks: \begin{equation} \mu:=\frac{m_um_d}{m_u+m_d}\end{equation} and noting the $c_3$ term vanishes for the case of two light quark flavours. The first term can be eliminated by making the field redefinition $\mathcal{N}\to e^{-i\alpha\gamma^5}\mathcal{N}$ which does generate some extra terms, however, they are not linear in the quark masses and can hence be neglected. Thus, only the second term contributes providing a pion-nucleon coupling that violates CP-symmetry. The value of $c_+$ is fixed by the proton neutron mass difference via: $c_+(m_u-m_d)=m_p-m_n=-1.3$MeV. Using $m_u=1.7$MeV and $m_d=3.9$MeV yields $c_+=0.6$.\\

We are finally in a position to write down, to leading order, the Lagrangian for a low-energy QCD EFT describing nucleons interacting via pions: 

\begin{equation}\label{EFT Lagrangian}\begin{aligned} \Lie&=  \frac{1}{2}\p^\mu\pi^a\p_\mu\pi^a-\frac{1}{6f^2_\pi}(\p^\mu\pi^a\p_\mu\pi^a\pi^b\pi^b-\p^\mu\pi^a\pi^a\p_\mu\pi^b\pi^b)\\
&-\frac{v^3}{f_\pi}(m_u+m_d)\left[\pi^+\pi^++\pi^-\pi^-\right]  -\frac{v^3}{f_\pi}(m_u+m_d)\left[1+\bar{\theta}^2\frac{(m_u-m_d)^2}{(m_u+m_d)^2}\right]\pi^0\pi^0\\
&+i \overline{\mathcal{N}} \slashed{\partial} \mathcal{N}-m_{N} \overline{\mathcal{N}} \mathcal{N}  -ig_{\pi\bar{N}N}\pi^a\overline{\mathcal{N}} \sigma^{a} \gamma_{5} \mathcal{N} -\frac{\bar{\theta} c_{+} \mu}{f_{\pi}} \pi^{a} \overline{\mathcal{N}} \sigma^{a} \mathcal{N}.\end{aligned}\end{equation}

\subsection{The Neutron eDM}

To calculate the eDM of the neutron, we need to evaluate a neutron-photon interaction. We can start by adding a term to the effective Lagrangian that represents the neutron eDM. Such a term can easily be written down: \begin{equation} \mathcal{L} \supset d_{n} F_{\mu \nu} \bar{n} S^{\mu \nu} i \gamma_{5} n\end{equation} where $S^{\mu\nu}:=\frac{i}{4}[\gamma^\mu,\gamma^\nu]$, $F^{\mu\nu}:=\p^\mu A^\nu-\p^\nu A^\mu$ is the electromagnetic field strength tensor, and the coupling constant of the photon-neutron interaction, $d_n$, is the neutron eDM. The corresponding Feynman rule is easily read off and we can thus write down the matrix element for photon-neutron scattering: \begin{equation}\label{first eDM} i\mathcal{M}=2 d_{n} \epsilon_{\mu}^{*}(q) \bar{u}\left(p^{\prime}\right) S^{\mu \nu} q_{\nu} i \gamma_{5} u(p).\end{equation}\\

\begin{figure}[h!]
\centering
\includegraphics[width=0.7\linewidth]{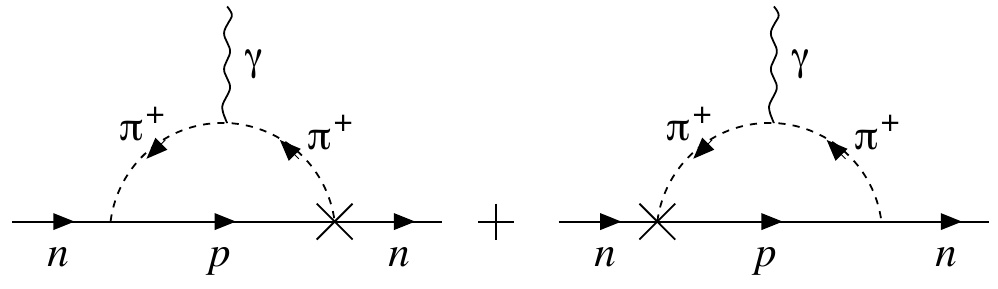}
\caption{The Feynman diagrams giving the leading-order contribution to the neutron eDM. The CP violating vertex is denoted with a cross \cite{QFT}.}
\label{eDM diagram}
\end{figure}

\begin{figure}[h!]
\centering
\includegraphics[width=0.35\linewidth]{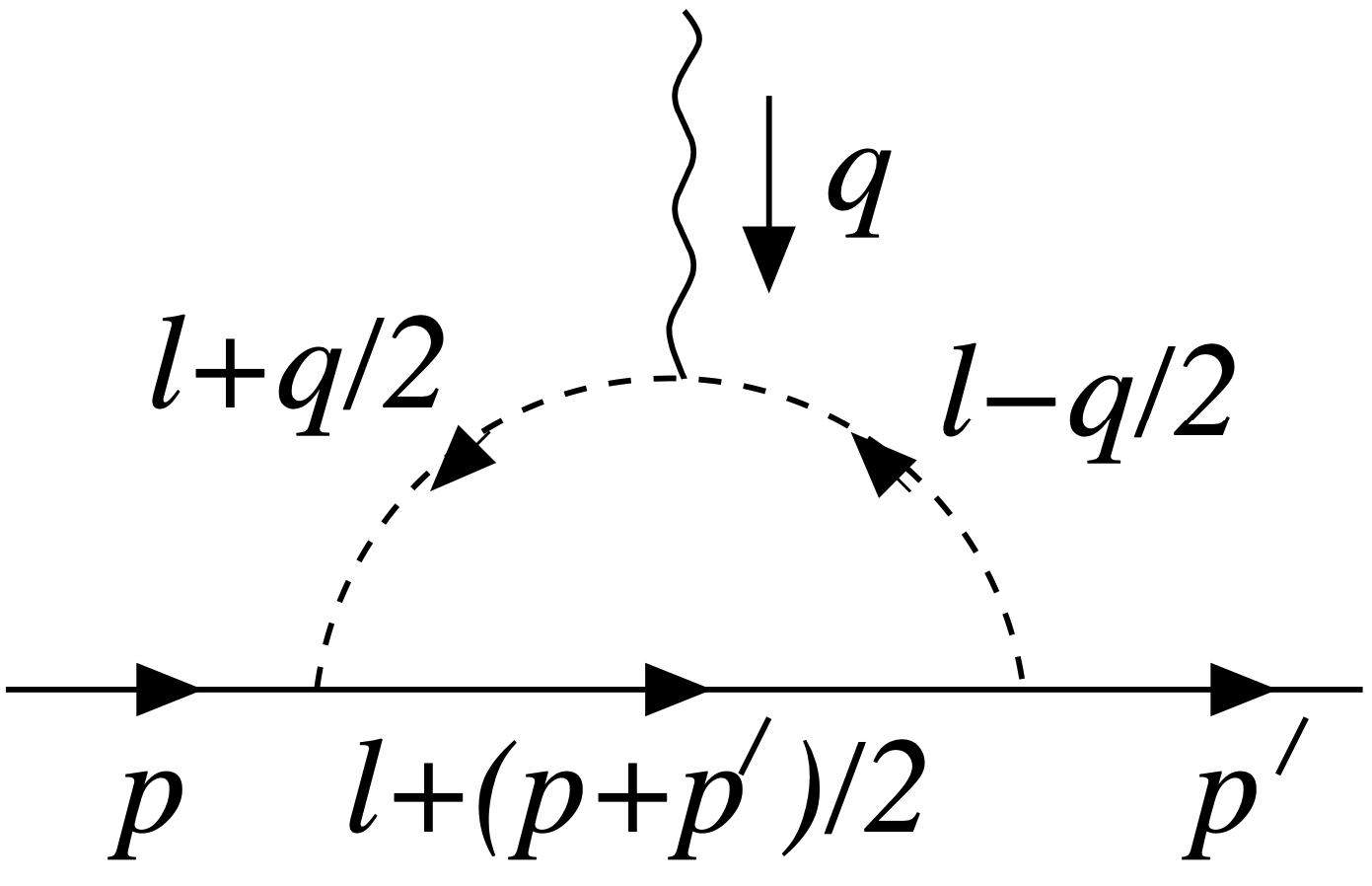}
\caption{The momentum flow in the diagrams of figure \ref{eDM diagram} \cite{QFT}.}
\label{momentum}
\end{figure}

We can now recalculate the matrix element for photon-neutron scattering in our low-energy QCD EFT (as developed in \S\ref{Low-Energy QCD}) and equate the two results to determine the neutron eDM. The leading order Feynman diagrams for photon-neutron scattering in our low-energy QCD EFT are shown in figure \ref{eDM diagram} with the corresponding momentum flow shown in figure \ref{momentum}. The reason for their leading order nature is due to an infrared divergence for small loop momenta since the pions have a low mass in comparison to the other available mass scales in the theory. This allows us to make the approximation $l\ll p,p'$, which will prove useful when calculating the matrix element. Note also that $l\ll p,p'$ means the internal proton is nearly on-shell and hence justifies our use of the Dirac equation in the fourth line of (\ref{piNN}), as promised.\\

We now evaluate the diagrams in figure \ref{eDM diagram} to calculate the photon-neutron scattering matrix element in our low-energy QCD EFT. The two relevant interaction terms from the effective Lagrangian are: \begin{equation} \Lie\supset -ig_{\pi\bar{N}N}\pi^a\overline{\mathcal{N}} \sigma^{a} \gamma_{5} \mathcal{N} -\frac{\bar{\theta} c_{+} \mu}{f_{\pi}} \pi^{a} \overline{\mathcal{N}} \sigma^{a} \mathcal{N}.\end{equation} Substituting the explicit forms for the nucleons (\ref{nucleon doublet}), the pions (\ref{explicit pions}), and the Pauli spin matrices yields: \begin{equation}\mathcal{L}\supset-i \sqrt{2}g_{\pi\bar{N}N}\left(\pi^{+} \bar{p} \gamma_{5} n+\pi^{-} \bar{n} \gamma_{5} p\right)-\sqrt{2}\frac{\bar{\theta} c_{+} \mu}{f_{\pi}}\left(\pi^{+} \bar{p} n+\pi^{-} \bar{n} p\right)\end{equation} and we can easily read off the Feynman rules for the CP violating and non-CP violating pion-nucleon vertices. With these results, along with the reference formulae in appendix \ref{A}, we can list all the Feynman rules we will require:

\begin{itemize}
\item non-CP violating $np\pi^+$ vertex: $\sqrt{2}g_{\pi\bar{N}N}\gamma_{5}$
\item CP violating $np\pi^+$ vertex: $i\sqrt{2}\bar{\theta} c_{+} \mu / f_{\pi}$
\item $\pi^+$ propagator: $\frac{i}{k^2-m_\pi^2}$,\quad $k=l\pm\frac{1}{2}q\quad$ (read off from the first term of (\ref{EFT Lagrangian}))
\item $\gamma\pi^+\pi^+$ vertex: $-ie(k_1+k_2)^\mu,\quad k_1=l-\frac{1}{2}q,\; k_2=l+\frac{1}{2}q\quad$ (Appendix \ref{SQED})
\item Nucleon propagator: $i\frac{\slashed{k}+m_N}{k^2-m_N^2},\quad k=l+\frac{1}{2}(p+p')\quad$ (Appendix \ref{QED})
\item Incoming fermion: $u(k),\quad k=p\quad$ (Appendix \ref{QED})
\item Outgoing fermion $\bar{u}(k), \quad k=p'\quad$ (Appendix \ref{QED})
\item Incoming photon: $\epsilon^*_\mu(k),\quad k=q\quad$ (Appendix \ref{QED})\\
\end{itemize}

We can now evaluate the Feynman diagrams in figure \ref{eDM diagram} with the above Feynman rules to obtain the neutron eDM. The matrix element reads: \begin{equation} \label{eDM Feynman}\begin{aligned}i \mathcal{M} &=-i e\sqrt{2}g_{\pi\bar{N}N} \frac{\sqrt{2}\bar{\theta} c_{+} \mu}{f_{\pi}} \epsilon_{\mu}^{*}(q)\\
&\times\int_0^\Lambda \frac{d^{4} l}{(2 \pi)^{4}}\frac{2 l^{\mu} \bar{u}\left(p^{\prime}\right)\left[\left(-\slashed{l}-\slashed{\bar{p}}+m_{N}\right) \gamma_{5}+\gamma_{5}\left(-\slashed{l}-\slashed{\bar{p}}+m_{N}\right)\right]u(p)}{(\left(l+\bar{p}\right)^{2}-m_{N}^{2})\left((l+\frac{1}{2}q)^{2}-m_{\pi}^{2}\right)\left((l-\frac{1}{2}q)^{2}-m_{\pi}^{2}\right)}\end{aligned}\end{equation} where $\bar{p}^\mu:=\frac{1}{2}(p'^\mu+p^\mu)$ and $\Lambda=4\pi f_\pi$ is the UV cutoff of our EFT. Using $\{\gamma^\mu,\gamma^5\}=0$, the numerator greatly simplifies to \begin{equation} \bar{u}(p')\left[\left(-\slashed{l}-\slashed{\bar{p}}+m_{N}\right) \gamma_{5}+\gamma_{5}\left(-\slashed{l}-\slashed{\bar{p}}+m_{N}\right)\right] u(p)=2 m_{N} \bar{u}(p') \gamma_{5} u(p).\end{equation} Using $\bar{u}(p') \gamma_{5} u(p)=0$ for $p'=p$ and noting that $q=p'-p$ tells us that $\bar{u}(p') \gamma_{5} u(p)$ is zero when $q$ is zero and therefore must be linear in $q$; we can thus set $q=0$ everywhere else. Using the aforementioned approximation $l\ll p,p'$ we can set $(l+\bar{p})^{2}-m_{N}^{2}=2 p \cdot l$ in the denominator. Substituting these simplifications brings our matrix element to \begin{equation} i\mathcal{M}=-i4e\frac{g_{\pi\bar{N}N}\bar{\theta} c_{+} \mu m_{N}}{f_{\pi}} \epsilon_{\mu}^{*}(q) \int_{0}^{\Lambda} \frac{d^{4} l}{(2 \pi)^{4}} \frac{\left(2 l^{\mu}\right) \bar{u}(p') \gamma_{5} u(p)}{(2 p \cdot l)\left(l^{2}+m_{\pi}^{2}\right)^{2}}.\end{equation} Any Lorentz vectors $l^\mu$ will be integrated out, but we must conserve the Lorentz structure. Since $p$ is the only other Lorentz vector present, it must gain the Lorentz index after the integral is performed. We can thus transfer the Lorentz indices from $l$ to $p$ before integrating and make the replacement: \begin{equation}\frac{l^{\mu}}{p \cdot l}=\frac{l^{\mu}p}{p_\nu l^\nu p} \rightarrow \frac{lp^{\mu}}{p_\nu lp^\nu}=\frac{p^{\mu}}{m_{N}^{2}}\end{equation} further simplifying our matrix element to \begin{equation} i\mathcal{M}=-i4e\frac{\bar{\theta} \lambda c_{+} \mu}{ f_{\pi}^{2}} \epsilon_{\mu}^{*}(q)p^\mu \bar{u}(p') \gamma_{5} u(p) \int_{0}^{\Lambda} \frac{d^{4} l}{(2 \pi)^{4}} \frac{1}{\left(l^{2}+m_{\pi}^{2}\right)^{2}}.\end{equation} Making use of the Gordon identity:\begin{equation} p^{\mu} \bar{u}(p') \gamma_{5} u(p)=\bar{u}(p') S^{\mu \nu} q_{\nu} i \gamma_{5} u(p)+O\left(q^{2}\right)\end{equation} verifies that $\bar{u}(p') \gamma_{5} u(p)$ is indeed linear in $q$ and simplifies our matrix element to \begin{equation} i\mathcal{M}=-i4e\frac{\bar{\theta} \lambda c_{+} \mu}{ f_{\pi}^{2}} \epsilon_{\mu}^{*}(q) \bar{u}(p') S^{\mu \nu} q_{\nu} i \gamma_{5} u(p) \int_{0}^{\Lambda} \frac{d^{4} l}{(2 \pi)^{4}} \frac{1}{\left(l^{2}-m_{\pi}^{2}\right)^{2}}.\end{equation} This is now written in terms of a standard integral which evaluates to $\left(i / 16 \pi^{2}\right) \ln \left(\Lambda^{2} / m_{\pi}^{2}\right)$. Our matrix element thus has the final form: \begin{equation}\label{eDM matrix element} i\mathcal{M}=e\frac{\bar{\theta} \lambda c_{+} \mu}{ 4\pi^2f_{\pi}^{2}} \epsilon_{\mu}^{*}(q) \bar{u}(p') S^{\mu \nu} q_{\nu}i\gamma_{5} u(p)  \ln \left(\frac{\Lambda^{2}}{m_{\pi}^{2}}\right).\end{equation}\\

By equating (\ref{eDM matrix element}) with (\ref{first eDM}), we can read off the neutron eDM: \begin{equation} d_{n}=\frac{e \bar{\theta} \lambda c_{+} \mu}{8 \pi^{2} f_{\pi}^{2}}\ln \left(\frac{\Lambda^{2}}{m_{\pi}^{2}}\right)\end{equation} and upon substituting the numerical values of the constants $(\lambda=1.27, \;c_{+}=0.6, \;\mu=1.2$ MeV, $\Lambda=4\pi f_\pi$, $f_\pi=130$ MeV, $m_\pi=135$ MeV) gives us a final calculated value of the neutron eDM: \begin{equation}\label{Calculated eDM} d_{n}=6.7 \times 10^{-17} \bar{\theta} e \text{ cm}.\end{equation} using the conversion eV$^{-1}=1.97\times 10^{-7}$ m.\\

We once again arrive at a physical quantity parameterised by $\bar{\theta}$, which is okay since $\bar{\theta}$ is invariant under the spurious symmetries (\ref{anom sym}). However, recalling the measured upper bound of the neutron eDM (\ref{eDM}), yields a constraint on $\bar{\theta}$ of \begin{equation} \label{theta constraint} \bar{\theta} \lesssim 10^{-10}\end{equation} and we have once again arrived at the strong CP problem at the quantum level. Recall $\bar{\theta}$ is comprised of the coefficient of the CP-violating term in the Lagrangian and the totally arbitrary phases of the up and down quark masses; there is no good reason for it to be this small. This truly is a problem.

\subsection{The QCD Axion}

One of the simplest solutions to the strong CP problem at the quantum level is what we call the \ti{axion EFT} (otherwise known as Peccei–Quinn theory). The EFT consists of a single new particle and coupling with an associated Lagrangian: \begin{equation}\label{PQ}\begin{aligned} \Lie_\text{PQ}=& \frac{1}{2} \partial_{\mu} a \partial^{\mu} a-\frac{1}{2} m_{\mathrm{a}}^{2} a^{2}+i \bar{u} \slashed{D} u-m_u\bar{u}e^{i\theta_u\gamma^5}u+i \bar{d} \slashed{D} d-m_d\bar{d}e^{i\theta_d\gamma^5}d\\
&-\frac{1}{4} F^{\mu \nu,a} F_{\mu \nu}^{a}-\left(\theta+\frac{a}{f_{a}}\right)\frac{g^{2}}{32 \pi^{2}} \tilde{F}^{\mu \nu,a} F_{\mu \nu}^{a}\end{aligned}\end{equation} where $a$ is the \ti{axion field}, $m_a$ is the mass of its quanta, and $f_a^{-1}$ is the axion-gluon coupling. Along with the addition of the axion kinetic and mass terms, this EFT amounts to making a replacement in the QCD Lagrangian (\ref{2 flavour L}) of \begin{equation}\label{replacement} \theta\to\theta+\frac{a}{f_{a}}.\end{equation} The axion vev, $\langle a\rangle$, is thus found from the value of the axion filed at the minimum of the potential (\ref{EFT mass term 2}) with the replacement (\ref{replacement}) applied; we simply need to solve \begin{equation} 2\text{Re}\left[\frac{\p V_\text{PQ}(a)}{\p {a}}\mid_{a=\langle a\rangle}\right]=0\end{equation} where \begin{equation}\begin{aligned} V_\text{PQ}(a)&=-v^3\text{Tr}\left[Me^{\frac{i\pi^a(x)\sigma^{a}}{f_\pi}}e^{i\left(\theta+\frac{a}{f_a}\right)}\right]\\
&=-v^3\text{Tr}\left[\begin{pmatrix}m_u&0\\0&m_d\end{pmatrix}e^{\frac{i\pi^a(x)\sigma^{a}}{f_\pi}}\right]e^{i\left(\bar{\theta}+\frac{a}{f_a}\right)}\\
&=-v^3\text{Tr}\left[\begin{pmatrix}m_u&0\\0&m_d\end{pmatrix}e^{\frac{i\pi^a(x)\sigma^{a}}{f_\pi}}\right]\left(1-i(\bar{\theta}+\frac{a}{f_a})-(\bar{\theta}+\frac{a}{f_a})^2\right)\end{aligned}\end{equation} to leading order in $\bar{\theta}$. This yields: \begin{equation}2\text{Re}\left[\frac{\p V_\text{PQ}(a)}{\p {a}}\mid_{a=\langle a\rangle}\right]=-2v^3\text{Re}\left[\text{Tr}\left[\begin{pmatrix}m_u&0\\0&m_d\end{pmatrix}e^{\frac{i\pi^a(x)\sigma^{a}}{f_\pi}}\right]\left(-\frac{i}{f_a}-\frac{2}{f_a}(\bar{\theta}+\frac{\langle a\rangle}{f_a})\right)\right]=0\end{equation} and thus the axion has vev: \begin{equation}\langle a\rangle=-\bar{\theta} f_a.\end{equation}\\

Recall our two offenders, the inconsistency of the $\pi_0$ mass and the larger than measured value of the neutron eDM. In the axion EFT, they each take the form: \begin{equation}\begin{aligned}\label{PQ results}
m_{\pi^0}^2&=2\frac{v^3}{f_\pi}(m_u+m_d)\left[1+(\bar{\theta}+\frac{a}{f_a})^2\frac{(m_u-m_d)^2}{(m_u+m_d)^2}\right]\\
d_{n} &= 3.2 \times 10^{16}\left(\bar{\theta}+\frac{a}{f_a}\right) \mathrm{e} \;\mathrm{cm}\end{aligned}\end{equation} by making the replacement (\ref{replacement}). When the axion sits in the minimum of its potential ($a\to\langle a \rangle=-\bar{\theta} f_a$), it dynamically restores the expected mass of the $\pi_0$ and sets the neutron eDM to zero, solving the strong CP problem in all its manifestations. This is exactly the same mechanism of dynamically minimising the potential that we saw in the classical axion solution to the strong CP problem, albeit taking far longer.\\

We finally made it - the strong CP problem is solved.

\subsection{Axion-Like Particles}

Now convinced that axions are an elegant solution to the strong CP problem we need to determine their properties. We start by finding the mass of the axion to leading order in $f^{-1}_\pi$ and $\bar{\theta}$: \begin{equation}\begin{aligned} m_{a}^2&=2\text{Re}\left[\frac{\p^2 V_\text{PQ}(a)}{\p {a}^2}\mid_{a=\langle a\rangle}\right]\\
&=-2v^3\text{Re}\left[\text{Tr}\left[\begin{pmatrix}m_u&0\\0&m_d\end{pmatrix}e^{\frac{i\pi^a(x)\sigma^{a}}{f_\pi}}\right]\frac{\p^2}{\p {a}^2}\left(1-i(\bar{\theta}+\frac{a}{f_a})-(\bar{\theta}+\frac{a}{f_a})^2\right)\mid_{\pi^a=\langle\pi^a\rangle}\right]\\
&=-2v^3\text{Re}\left[\left(m_u+m_d\right)\left(-\frac{2}{f_a^2}\right)\right]\\ \end{aligned}\end{equation} and thus we find the axion-gluon coupling to be directly related to the axion mass: \begin{equation} \frac{1}{f_a}=\frac{m_a}{2v^\frac{3}{2}\sqrt{m_u+m_d}}.\end{equation} We can eliminate $v^3$ by substituting $m_{\pi^0}$ (\ref{PQ results}): \begin{equation} \frac{1}{f_a}=\frac{\sqrt{2}}{2m_{\pi^0}f_\pi}m_a\end{equation} and inputting the values $f_\pi\approx 130$ MeV and $m_{\pi^0}\approx 135$ MeV yields: \begin{equation} \frac{1}{f_a}\approx 4\times10^{-17} \text{eV}^{-2} m_a,\end{equation} clearly extremely small in comparison to the mass of the axion.\\

\begin{figure}[h!]
\centering
\includegraphics[width=0.9\linewidth]{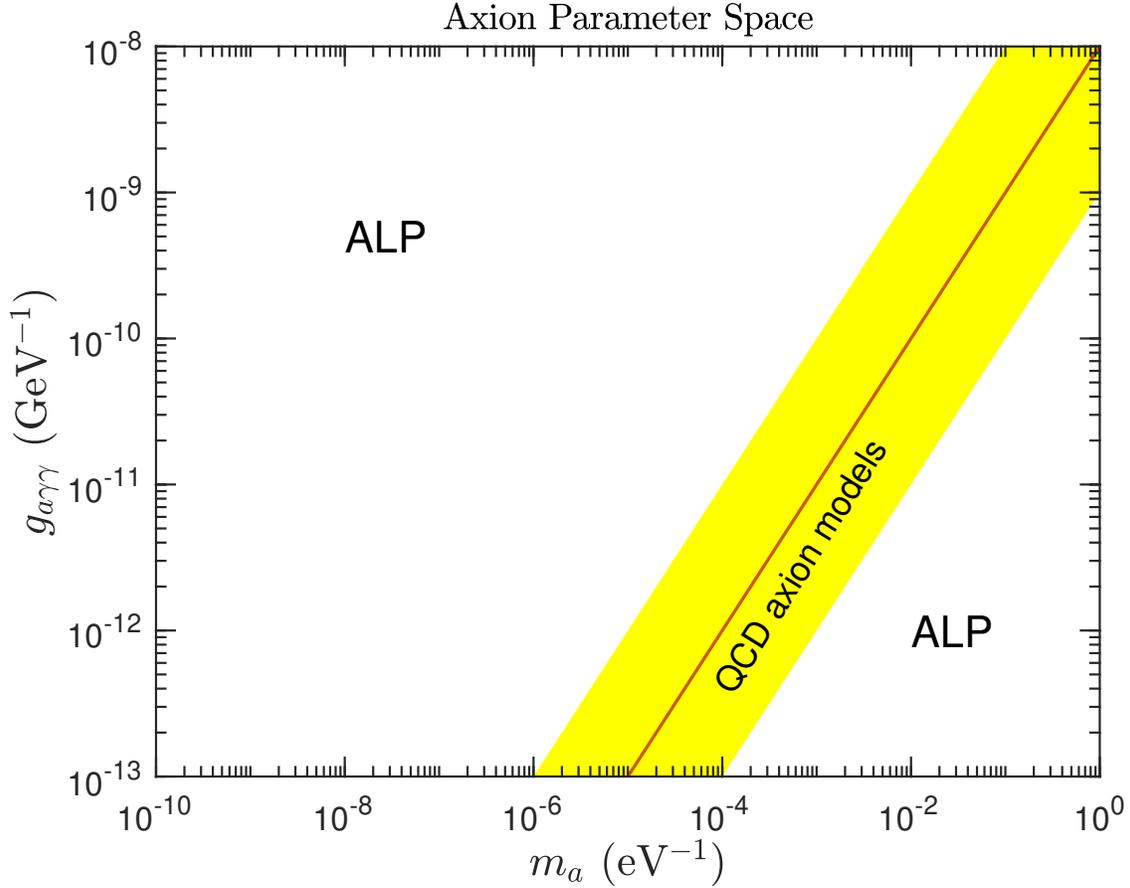}
\caption{The axion parameter space. The red line is a plot of (\ref{gayy}) and the yellow region displays the band of QCD axion models.}
\label{Parameter space}
\end{figure}

So far we have explored the axion-gluon coupling which naturally arises when constructing the axion EFT to solve the strong CP problem. Having introduced a new particle we should ask the question: what else could the axion couple to? It is reasonable to postulate an axion-photon coupling which is of the same order as our axion-gluon coupling: \begin{equation} \label{gayy} g_{a\gamma\gamma}\sim10^{-17} \text{eV}^{-2} m_a\end{equation} where $g_{a\gamma\gamma}$ is the axion-photon coupling. If this is indeed the case, the direct link between the mass of the axion and its coupling to photons, alongside a coupling to photons being easy to probe, makes this a popular method for experimental searches of axions. The region in axion parameter space around this linear relationship is known as the \ti{band of QCD axion models} and is where we could expect to detect a QCD axion; this is displayed by the yellow region in figure \ref{Parameter space} where the red line is the linear relationship (\ref{gayy}).\\

However, we can go one further. The axion solves the strong CP problem via its coupling to the gluons and thus a coupling to photons has absolutely nothing to do with the strong CP problem. So the question is: if we are searching for an axion via its coupling to photons, why does it have to solve the strong CP problem? Well, it doesn't. We call these particles \ti{axion-like particles} (ALP). Recall the linear relationship between the axion coupling and its mass comes from its requirement to solve the strong CP problem, but since an ALP is not required to solve the strong CP problem no such relationship is required. The connection between their coupling to photons and their mass is completely severed; they are not restricted to the band of QCD axion models and could lie \ti{anywhere} in the axion parameter space. Thus, when experimentally searching for the axion we do not need to restrict ourselves to the band of QCD axion models and can probe all of the axion parameter space, for the detection of an ALP would in itself be strong evidence for the existence of the QCD axion.\\

This all seems a bit abstract, and that's because it is. Perhaps the best motivation for all of this is that with ever-improving detectors we can probe regions of axion parameter space far greater than just the band of QCD axion models, so we should look and see what we find.\\

This naming convention of QCD axions and ALPs can be a slightly confusing one. We assure the reader that any previously mentioned `axion' is indeed a QCD axion, however, from here onwards an `axion' could refer to either a QCD axion, an ALP, or both and is determined from the context. To summarise:

\begin{itemize}
\item QCD Axion: Solves the Strong CP problem
\item ALP: Does not solve the Strong CP problem
\item Axion: Either or both of the above, determined by the content.
\end{itemize}

\newpage

\section{Experimental Searches For Axions}

\subsection{Axion-Photon Conversion}

To use the axion-photon coupling as a means of detecting axions, we first need to determine the probability of axion photon conversion. The derivation itself provides little enlightenment towards our goal so we demote it to appendix \ref{APP} and simply state the result: \begin{equation}\label{conversion prob 2} P(\gamma\to a)=P(a\to\gamma)=\left(\frac{g_{a \gamma} B_{e}}{q}\right)^{2} \sin ^{2}\left(\frac{q L}{2}\right).\end{equation} The above applies for an axion propagating along an optical cavity of length $L$ filled with an external magnetic field of strength $B_e$, where $q$ is the difference between the axion and photon wavenumbers.\\

The key point is that axion-photon conversion is stimulated by the presence of an external magnetic field $B_e$. Thus, to best detect axions via this coupling we must construct a long optical cavity filled with a strong magnetic field where axion-photon conversion can occur and the resulting photons can be detected.

\subsection{Solar Axion Production} 

\begin{figure}[h!]
\centering
\includegraphics[width=0.6\linewidth]{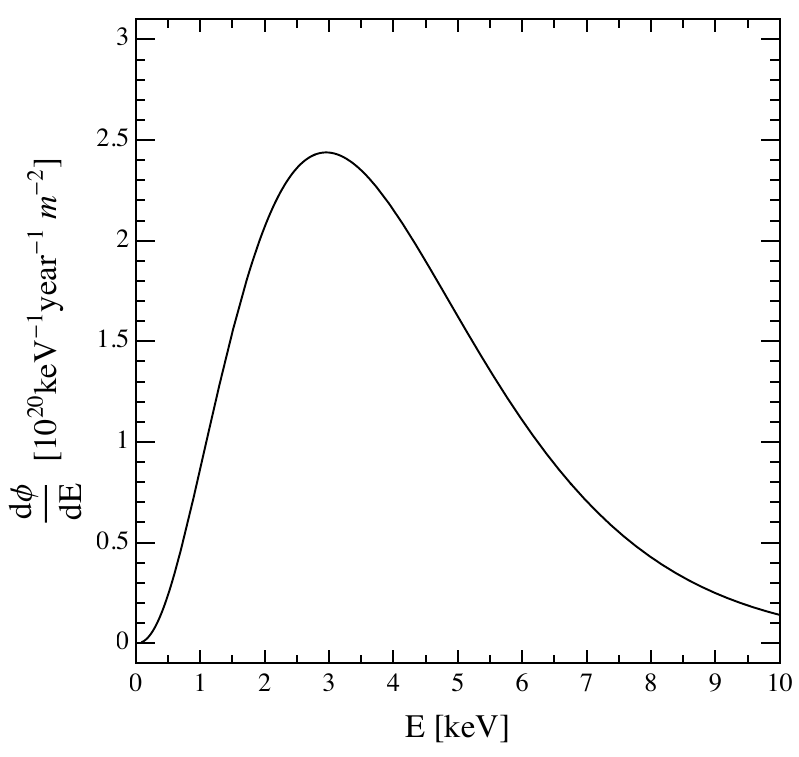}
\caption{Solar axion flux spectra at earth assuming only the Primakoff conversion of plasma photons into axions. An illustrative value $g_{a\gamma\gamma}=10^{-12}$GeV$^{-1}$ of the coupling constant has been chosen \cite{Exp}.}
\label{Prim}
\end{figure}

Axions can be produced within the solar interior via a process called the \ti{Primakoff conversion}, where plasma photons are converted into axions due to the presence of the Coulomb field of charged particles. Although there are other production methods (such as ABC mechanisms where the ALPs couple with electrons) the detection methods we study only consider the Primakoff conversion channel. This channel peaks at 4.2 keV and exponentially decreases for higher energies, as seen in figure \ref{Prim}.

\subsection{The Axion Helioscope}

%The helioscope was first used by Benedetto Castelli between 1578 and 1643 and later refined by Galileo between 1564 and 1642 \cite{Gal}. It was a device for observing the sun and its sunspots through a projection of its image onto a white piece of paper with the use of a telescope. Its long distant cousin, \ti{the axion helioscope}, shares only about the same genes in that they both look at the sun.\\

\begin{figure}[h!]
\centering
\includegraphics[width=0.9\linewidth]{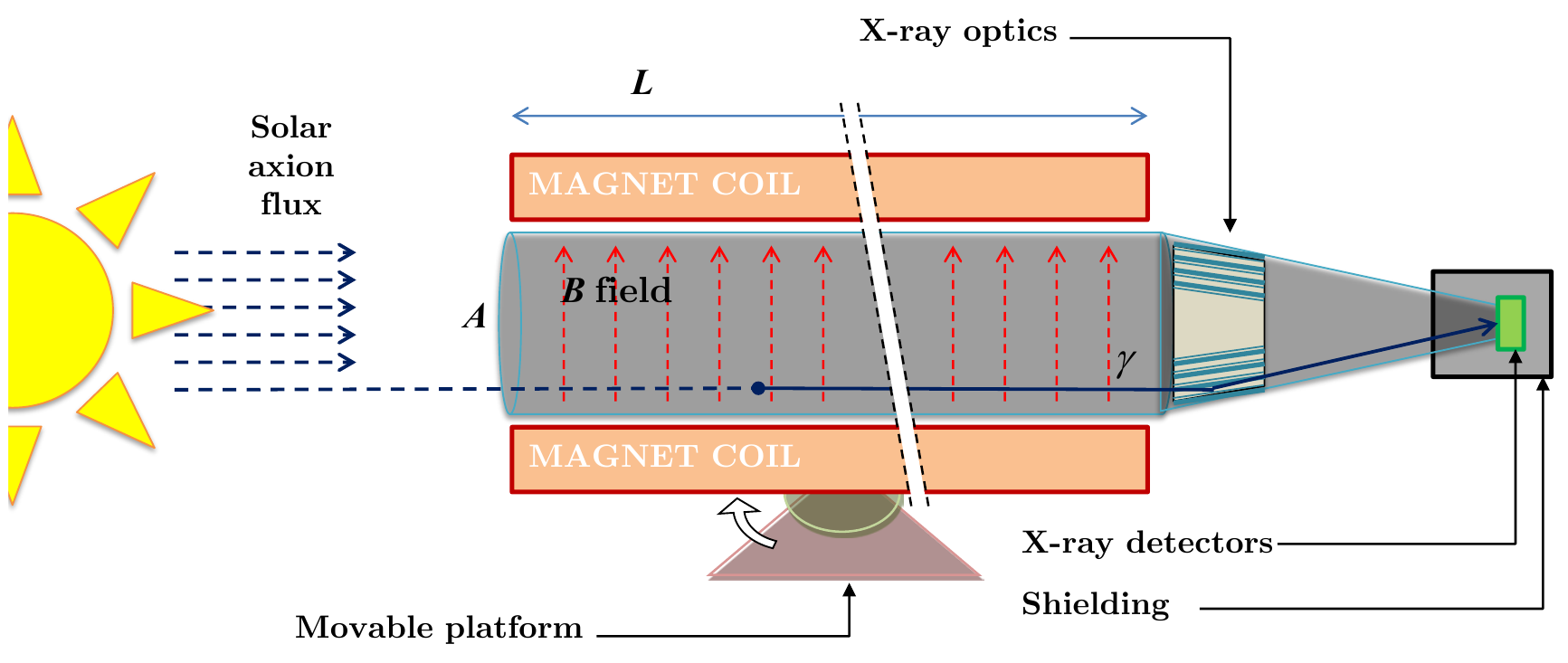}
\caption{Conceptual arrangement of an enhanced axion helioscope with X-ray focusing. Solar axions enter the optical cavity and are converted into photons due to the applied transverse magnetic field. The resulting wide beam of photons are focused into a narrow beam onto a detector \cite{Exp}.}
\label{helioscope}
\end{figure}

A popular tool for experimentally probing the axion is called the \ti{axion helioscope}. Using a source of solar axions, they stimulate axion-photon conversion by the means of a strong laboratory magnet and aim to detect the resulting X-rays produced. An axion helioscope will thus consist of a powerful magnet applying a strong magnetic field to a long optical cavity (the magnets bore), where axion-photon conversion can occur. This is combined with X-ray detectors to detect such a conversion, with an optional X-ray focusing stage between the magnet and the detector to increase the signal to noise ratio. See figure \ref{helioscope} for a conceptual arrangement of an enhanced axion helioscope with X-ray focusing.\\

Aligning the optical cavity with the sun will, hopefully, yield a spike in X-ray detection due to solar axion production. In the event of such a spike, (\ref{conversion prob 2}) can be used to determine $g_{a\gamma\gamma}$ and thus measure the mass of the axion, $m_a$. If no signal above the background is observed upon solar alignment of the axion helioscope, a notion we will become very familiar with, an experimental upper bound of the axion-photon coupling $g_{a\gamma\gamma}$ can be determined.\\

Axion helioscopes need only consider the Primakoff conversion channel as this maintains the broadest generality and produces relevant limits on $g_{a\gamma\gamma}$ over large mass ranges. For a static background field, the energy of the reconverted photon is identical to that of the incoming axion. We thus expect to detect the same photon energy distribution as seen for the axions in figure \ref{Prim}, with the same peak at $4.2$ keV (X-rays). Coherent conversion along the whole length of the magnets bore occurs when $qL\ll1$. For relativistic axions in vacuum, the difference between the axion and photon wavenumbers, $q$, is given approximately by $q=k_\gamma-k_a\approx \frac{m^2_a}{2\omega}$. The coherence condition is then satisfied, for the expected solar axion energies, with an optical cavity length of $L\sim 10$m, given the axion has mass: \begin{equation}\label{axion mass constraint}m_{a} \lesssim 10^{-2} \mathrm{eV}.\end{equation} With $(qL)^2\propto m_a^4$, the sensitivity of the experiment decreases $\sim m^{-4}_a$ for larger masses.\\

A buffer gas can be added to the optical cavity to increase sensitivity to higher mass axions. The gas imparts an effective mass of $m_\gamma=\omega_p$ to the photons, where $\omega_p$ is the plasma frequency of the gas given by: \begin{equation} \omega_p^2 =\frac{4\pi\alpha n_e}{m_e}\end{equation} with $n_e$ and $m_e$ denoting the electron number density and mass and $\alpha=\frac{1}{137}$ is the fine structure constant. If the axion mass matches the effective photon mass $(m_a=\omega_p)$ then $q=0$ and the coherence condition is restored, thus increasing the sensitivity to higher mass axions.\\

To help us build the most effective axion helioscope we define the \ti{figure of merit}, $f_M$, which characterises the effectiveness of axion-photon conversion of a helioscope's magnet. Thus, when designing an axion helioscope, maximising $f_M$ will be the main objective. The rate of axion-photon conversion is given by: \begin{equation} A\times P(a\to\gamma)\sim AB^2L^2\end{equation} and we thus define the \ti{figure of merit} as \begin{equation}\label{fm}f_M:=AB^2L^2.\end{equation}

\subsection{The Rise of the Axion Helioscope}

The first axion helioscope was achieved in 1992 at the Brookhaven National Laboratory (BNL), where a stationary dipole magnet with a field of $B = 2.2$ T and a length of $L = 1.8$ m was oriented towards the setting sun, hoping to detect a spike in X-rays as the sun passed over the aperture. No signal above the background was observed and thus the experiment set an upper limit of the axion photon coupling of $g_{a \gamma\gamma}<3.6 \times 10^{-9}$ GeV$^{-1}$ for an axion mass range of $m_{a}<0.03$ eV, and $g_{a \gamma\gamma}<7.7 \times 10^{-9}$ GeV$^{-1}$ for an axion mass range of $0.03\text{ eV}<m_{a}<0.11$ eV, both at $99 \%$ C.L. \cite{BNL}.\\

The second generation of axion helioscopes, SUMICO, was produced at the University of Tokyo, first achieving measurements by 1998. Improvements over the BNL axion helioscope include dynamic tracking of the sun, an evacuated optical cavity of length $L=2.3$ m, a far stronger applied magnetic field of $B=4$ T, and higher-performance X-ray detectors. SUMICO provided an upper limit of the axion photon coupling 4.5 times more stringent than the BNL helioscope of $g_{a \gamma\gamma}<6.0 \times 10^{-10} \mathrm{GeV}^{-1}$ for $m_{a}<0.03 \mathrm{eV}$ at $95 \%$ C.L upon detecting no signal above the background \cite{SUMICO1}.\\

In 2002, a buffer gas was added to the magnet's bore to increase sensitivity to higher mass axions. This allowed SUMICO to probe axions of mass $0.05$ eV $<m_a<0.27$ eV and upon detecting no signal above the background set an upper limit of $g_{a \gamma\gamma}<6.8-10.9 \times 10^{-10}$ GeV$^{-1}$ at 95\% C.L. \cite{SUMICO2}. By 2008, a higher mass range of $0.84$ eV $<m_a<1.00$ eV was probed and set an upper limit of $g_{a \gamma\gamma}<5.6-13.4 \times 10^{-10}$ GeV$^{-1}$ at 95\% C.L. upon detecting no signal above the background \cite{SUMICO3}.

\subsection{The Legacy of the CERN Axion Solar Telescope}

\begin{figure}[h!]
\centering
\includegraphics[width=\linewidth]{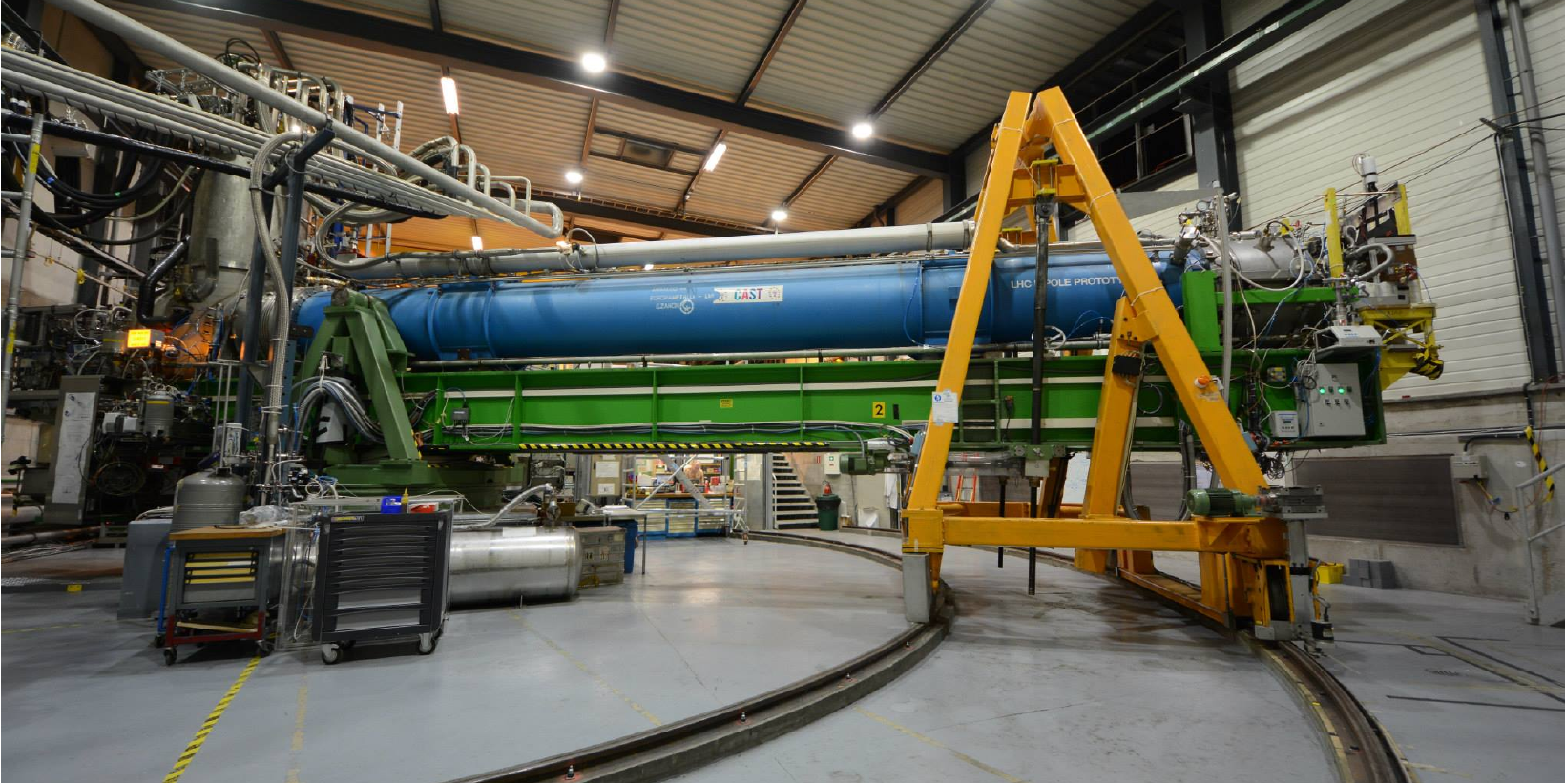}
\caption{Picture of the CAST experiment at CERN. Credit: M. Rosu/CAST collaboration, CERN.}
\label{CAST}
\end{figure}

The CERN Axion Solar Telescope (CAST) is a third-generation experiment, beginning data collection in 2003. CAST uses a decommissioned LHC test magnet of length 9.3 m and a magnetic field of up to 9 T \cite{Kuster}. The helioscope is equipped with an elevation and azimuth drive to track the sun over several hours each day and is the first helioscope to employ X-ray focusing optics and low background techniques, via the use of underground detectors. Figure \ref{CAST} shows a picture of the CAST experiment at CERN and a time-lapse of CAST tracking the sun throughout the day can be found on CERN's website \cite{CERN}.\\

In 2003, CAST operated for roughly 6 months with the magnet's bore in vacuum to probe a mass range of $m_a\lesssim0.2$ eV. No signal above the background was observed and thus CAST set a new upper limit on the axion-photon coupling of $g_{a\gamma\gamma\gamma}<1.16\times 10^{-10}$ GeV$^{-1}$ at 95\% C.L. \cite{CAST1}.\\ 

CAST was soon upgraded, operating between 2005 and 2006 with $^4$He contained in the optical cavity to increase its sensitivity to higher mass axions. Within this period, CAST operated at around 160 pressure settings taking approximately 2 hours of data at each setting. Once again, no signal above the background was observed and thus CAST set a new upper limit of $g_{a\gamma\gamma}<2.2\times 10^{-10}$ GeV$^{-1}$ at 95\% C.L. for a mass range of $m_a\lesssim 0.4$ eV \cite{CAST2}.\\

From 2008 to 2011, the $^4$He within the optical cavity was exchanged for $^3$He, allowing for higher pressure settings and hence sensitivity to higher mass axions. CAST first operated with the $^3$He gas at $T=1.8$ K, taking approximately 1 hour of data at each of the 252 different pressure settings, probing an axion mass range of 0.39 eV $\lesssim m_a\lesssim$ 0.64 eV. Once again, no signal above the background was observed and thus set a upper limit for this mass range of $g_{a \gamma\gamma} \lesssim 2.3 \times 10^{-10}$ GeV$^{-1}$ (95\% C.L.) \cite{CAST3}. CAST then went on to probe the mass range of 0.64 eV $<m_a<$ 1.17 eV and, as usual, no signal above the background was observed setting an upper limit for this mass range of $g_{a \gamma\gamma} \lesssim 3.3 \times 10^{-10}$ GeV$^{-1}$ (95\% C.L.) \cite{CAST4}.\\

In recent years (2013 to 2014), CAST revisited the vacuum phase, once again probing the mass range of $m_a\lesssim0.2$ eV. However, CAST now had the aid of improved detectors and novel X-ray optics, courtesy of R\&D for the next generation of axion helioscopes (IAXO), increasing the signal-to-noise ratio by a factor of 3 over CAST's previous operational periods. Unfortunately, perhaps to no surprise by now, no signal above the background was observed. However, this operational period was able to set a record upper limit on the axion-photon coupling of \cite{CAST5}: \begin{equation} g_{a \gamma\gamma}<0.66 \times 10^{-10}\text{ GeV}^{-1}\;(95\%\text { C.L.}).\end{equation}\\

Although all the expeditions of our helioscopes seem to be rather unfruitful, the upper limit of $g_{a\gamma\gamma}$ set by CAST's most recent adventure really is a profound achievement; for we are now probing deep into the  band of QCD axion models (\ref{gayy}) for low mass axions. For an axion mass of $m_a\lesssim0.2$ eV we currently have: \begin{equation}\left(\frac{g_{a\gamma\gamma}}{m_a}\right)_\text{QCD axion model}\lesssim 10^{-16},\quad\quad \left(\frac{g_{a\gamma\gamma}}{m_a}\right)_\text{Experimental} < 3.3\times 10^{-19}.\end{equation}

\subsection{The International Axion Observatory: The Final Frontier}

\begin{figure}[h!]
\centering
\includegraphics[width=0.5\linewidth]{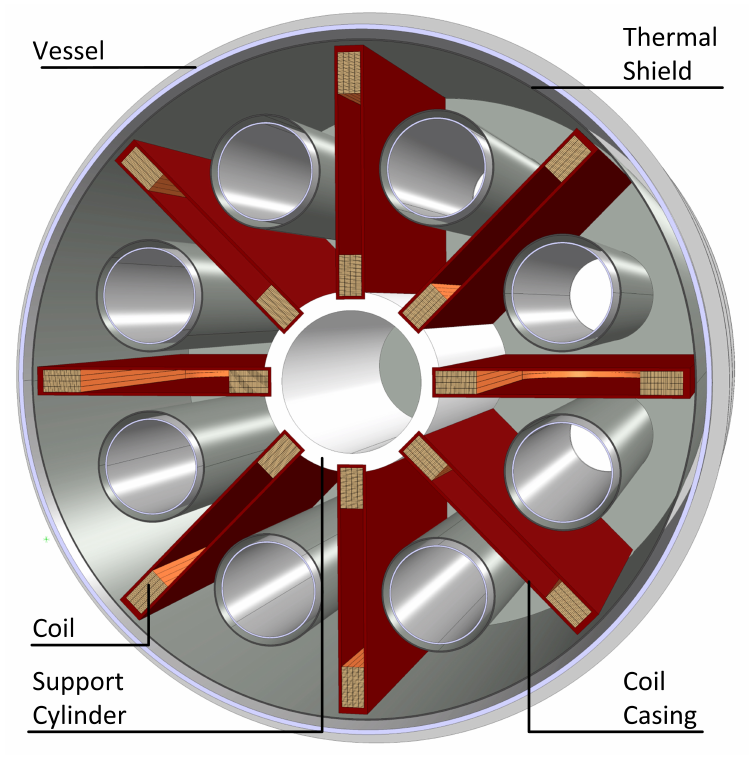}
\caption{Cross-section of IAXO's magnet and its bores in the area maximising arrangement \cite{Shilon}. The cold mass is surrounded by a thermal shield and a cryostat. The open bores allow for easy regular maintenance of the magnet system.}
\label{IAXO cross section}
\end{figure}

\begin{figure}[h!]
\centering
\includegraphics[width=0.7\linewidth]{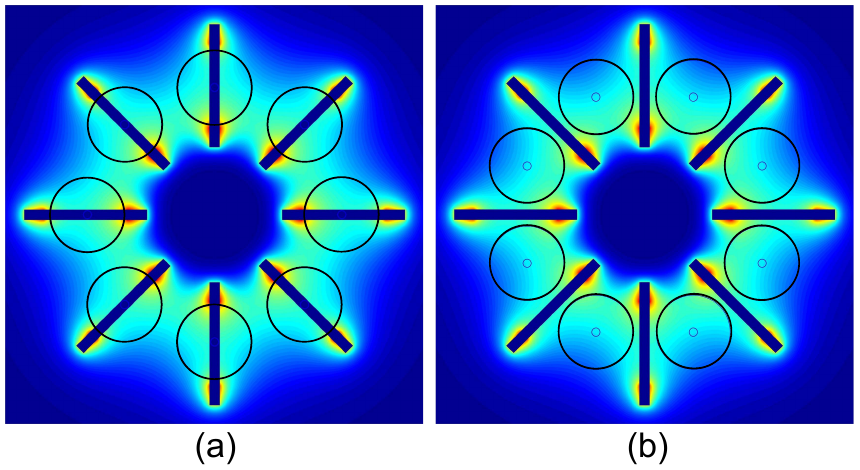}
\caption{Illustration of the two possible alignments of the bores and coils in IAXO \cite{Shilon}. The rectangles represent the toroid’s coil and the circles represent the telescopes’ bores. (a) Field maximising: Telescope bores are centred inside the superconducting coils. (b) Area maximising: Telescope bores are placed between the superconducting coils.}
\label{IAXO alignment}
\end{figure}

The International Axion Observatory (IAXO) is the next generation of axion helioscope, currently at the design stage. Its main asset is a new, purpose-built, superconducting magnet in a toroidal multibore configuration (see figure \ref{IAXO cross section}) which will efficiently produce an intense magnetic field over a large volume. Recall that, when designing an axion helioscope, our objective is to maximise the figure of merit $f_M$ (\ref{fm}), thus, there are two possible alignments of the bores and coils in IAXO. In the first, the bores are placed between the superconducting coils (area maximising) and in the second, the bores are centred inside the superconducting coils (field maximising). See figure \ref{IAXO alignment} for a pictorial description.\\

\begin{figure}[h!]
\centering
\includegraphics[width=0.7\linewidth]{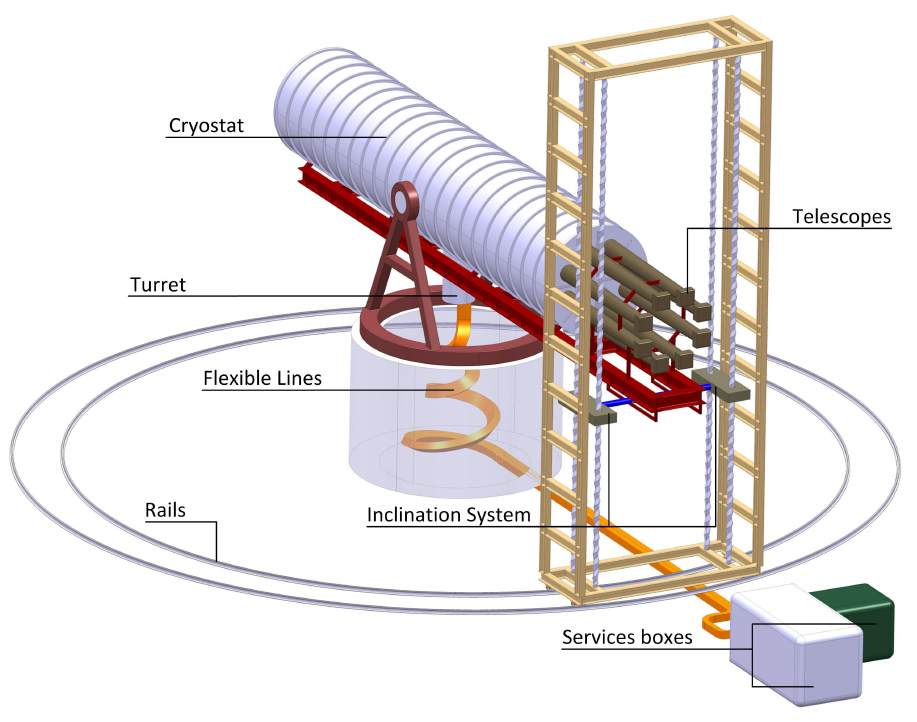}
\caption{Schematic view of IAXO. Shown are the cryostat, eight telescopes, turret leading the flexible services lines into the magnet, cryogenics and powering services boxes, inclination system, and the rails for horizontal movement \cite{Shilon}.}
\label{IAXO}
\end{figure}

The current design for the magnet is a 22 m long toroid with a 2.05 m outer radius and a 1.05 m inner radius, constructed from 8 superconducting coils generating a peak magnetic field of 5.4 T at the coil centres (relevant to the field maximising arrangement) and an average of 2.5 T in the 8 telescope bores of 0.6 m diameter (for the area maximising arrangement) \cite{Shilon}. The area maximising arrangement has the largest value of $f_M$ (300 times that of CAST) and is thus the configuration of choice. Each of the 8 bores is equipped with a 0.6 m diameter x-ray telescope and the project aims to use optimised optics based on thermally-formed glass substrates, similar to those used on NASA’s NuSTAR \cite{Harrison}. At the focal plane of each of the optics are low background pixelated detectors to image the focused signal. IAXO will also feature a buffer gas phase to increase its sensitivity to higher mass axions. Figure \ref{IAXO} shows the conceptual design of the overall infrastructure.\\

The aforementioned detectors are the exact ones used in the latest iteration of CAST and consist of small gaseous chambers read by pixelised planes of microbulk Micromesh Gas Structures (Micromegas). The latest generation of such detectors in CAST have achieved record-breaking background levels of $1.5\times10^{-6}$ counts keV$^{-1}$ cm$^{-2}$ s$^{-1}$, more than 100 times lower than the ones obtained by the first generation of CAST detectors \cite{Aune}. Values down to $\sim10^{-7}$ counts keV$^{-1}$ cm$^{-2}$ s$^{-1}$ have been obtained in a test bench placed underground in the Laboratorio Subterraneo de Canfranc (LSC) \cite{Irastorza}. Further reducing these values to $\sim10^{-8}$ counts keV$^{-1}$ cm$^{-2}$ s$^{-1}$ is currently being worked on, showing the good prospects of this technology for the future application in IAXO.\\

Putting all the discussed values together, IAXO is expected to improve the CAST signal-to-noise ratio by more than a factor $10^{-4}$, corresponding to setting a new upper limit on $g_{a\gamma\gamma}$ more than one order of magnitude smaller in the event of no signal above the background.\\

To begin the journey to IAXO, a scaled-down version of the experiment, BabyIAXO, is to be first constructed. BabyIAXO will use only two of the 8 modules present in IAXO (two coils, two bores, and two X-ray telescopes) but with a reduced length of 10 m. BabyIAXO will detect or reject axions with an axion-photon coupling down to $g_{a\gamma\gamma}\sim 1.5\times 10^{-11}$ GeV$^{-1}$ for axion masses up to $m_a\sim0.25$ eV. BabyIAXO will therefore constitute a prototype for the final infrastructure of IAXO, but also a fully-fledged axion helioscope, exceeding CAST's current limits \cite{Abeln}.

\subsection{A Summary of Axion Helioscopes} 

\begin{figure}[h!]
\centering
\includegraphics[width=0.9\linewidth]{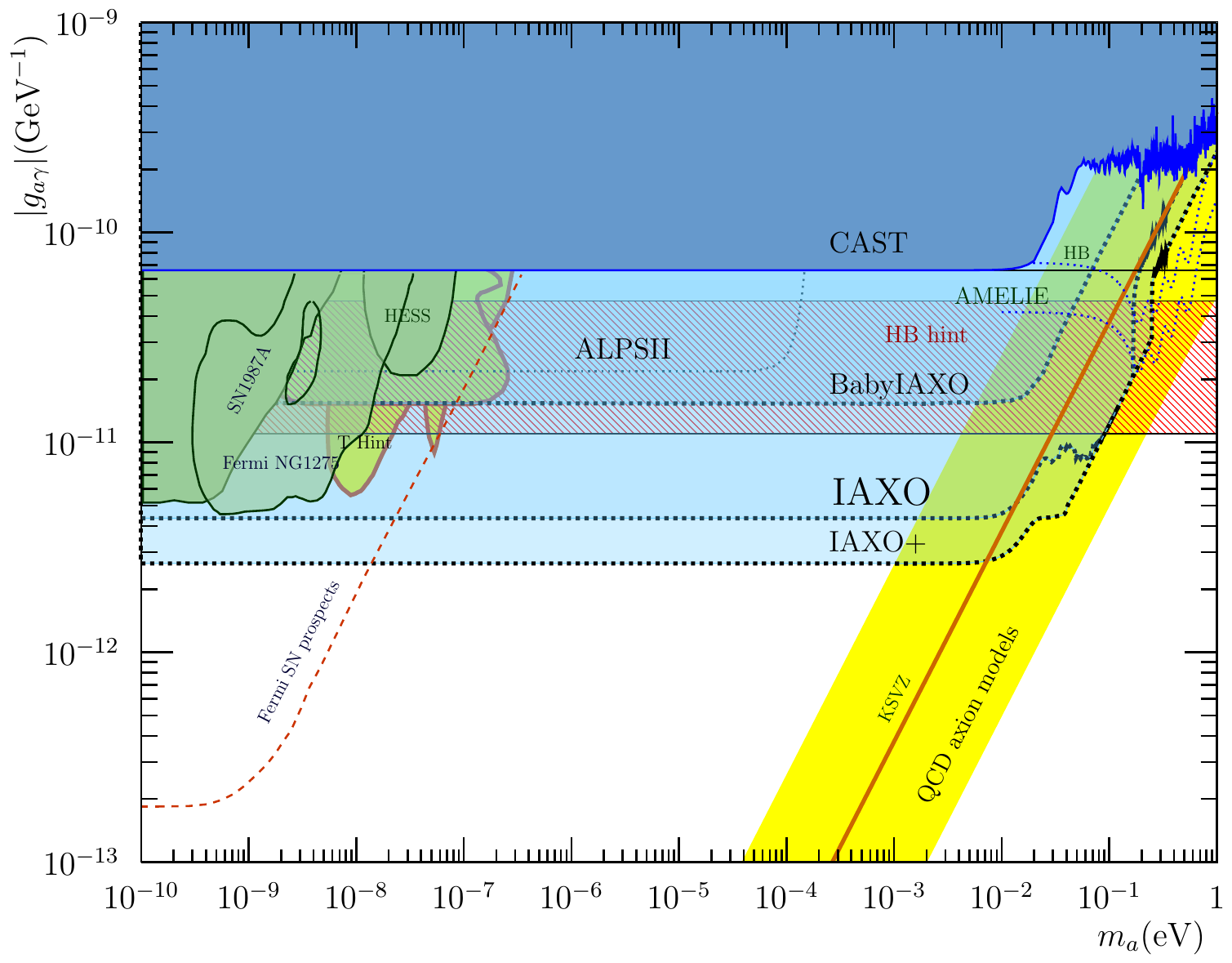}
\caption{The current (or expected) sensitivity of axion helioscopes in the axion parameter space. The yellow band displays the region of QCD axion models \cite{Exp}.}
\label{Helioscopes graph}
\end{figure}

We finally provide all of the details of the above three sections in table $\ref{Helioscopes table}$. Figure \ref{Helioscopes graph} displays the current (or expected) sensitivity of axion helioscopes in the axion parameter space. As one can see, the hunt for axions contains many other chapters with many other ongoing experiments, we just focused on the most relevant parts of the story. The red line is a plot of our justified QCD axion coupling to photons (\ref{gayy}) and thus the yellow region is considered the band of possible QCD axion models. CAST can be seen to probe deep into the QCD axion model band at its experimental limits, but only for light mass axions. Both BabyIAXO and IAXO will sufficiently probe a large portion of the QCD axion model band, imposing exciting prospects for future detection.

\begin{table}[h!]
\centering
\begin{tabular}{ccccccccc}
\toprule
Experiment&References&Status&$B$ (T)&$L$ (m)&$A$ (cm$^2$)& $f_{M}$ (Tm$^3$)&Optics&$\tilde{g}$\\
\midrule
BNL&\cite{BNL}&Past&2.2&1.8&130&0.20&No&36\\

SUMICO&\cite{SUMICO1,SUMICO2,SUMICO3}&Past&4&2.3&18&0.15&No&6\\

CAST&\cite{Kuster} - \cite{CAST5}&Ongoing&9&9.3&30&0.25&Yes&0.66\\

BabyIAXO&\cite{Abeln}&In design&$2.5$&10&$5.6\times10^3$&$350$&Yes&0.15\\

IAXO&\cite{Shilon} - \cite{Irastorza}&In design&$2.5$&22&$2.3\times10^4$&$7000$&Yes&0.04\\

\bottomrule
\end{tabular}
\caption{Past and future axion helioscopes with their key features. The last column represents the sensitivity achieved (or expected) in terms of an upper limit on $\tilde{g}=g_{a\gamma\gamma}\times10^{10}$ GeV for low mass axions ($m_a\lesssim0.2$ eV). The numbers for BabyIAXO and IAXO helioscopes correspond to design parameters from the quoted references.}
\label{Helioscopes table}
\end{table}

\newpage

\section{Epilogue}

We have achieved much in our journey. Recall it all began with our freedom to add a term to the QCD Lagrangian (parameterised by $\theta$) that allows for CP-violating interactions. Our first inkling of a problem is that no such CP-violating interactions have ever been observed in a strongly interacting experiment. Our first piece of evidence was the differing of the pion masses by the parameter $\bar{\theta}$, and since the pions are of roughly equal mass suggest that $\bar{\theta}$ must be small. Our second piece of evidence was the calculation of the neutron eDM, yielding a value (in terms of $\bar{\theta}$) that is far larger than we experimentally measure it to be, once again telling us that $\bar{\theta}$ must be small. These two pieces of evidence instruct us to add something into our theory that will dynamically set the parameter $\bar{\theta}$ to zero, a new particle called the \ti{axion}. We then discussed experimental probes of the axion, and although they seemed to be rather unfruitful, we will very soon be covering a huge section of axion parameter space and probe deep into the band of QCD axion models. It truly is an exciting time to be an axion.

\newpage

\begin{singlespace}

\begin{appendices}

\section{Reference formulae} \label{A}

%When applying Feynman rules, momentum is conserved at each vertex and any undetermined loop momenta are integrated over: multiply by a factor of $\int\frac{d^4l}{(2\pi)^4}$. Diagrams obtain relative minus signs due to spin statistics: a factor of $-1$ for each fermion loop and a relative factor of $-1$ for identical particle exchange. Finally, each diagram gains a topological symmetry factor where applicable. $\{\mu,\nu\}$ are Lorentz indices and $\{\alpha,\beta\}$ are spinor indices.

\subsection{Scalar QED Feynman Rules} \label{SQED}

\begin{minipage}{0.5\textwidth}

\begin{tikzpicture}\begin{feynman}[small]
\diagram{a [dot] -- [scalar, momentum=$p$, with arrow=0.5] b [dot] };
\end{feynman}\end{tikzpicture}
$=\tilde{S}(p)=\frac{i}{p^{2}-m^{2}+i \epsilon}$\\
\\
\begin{tikzpicture}\begin{feynman}[small]
\diagram{a [dot] -- [boson, momentum=$p$] b [dot] };
\vertex [above=0.6em of a] {$\mu$};
\vertex [above=0.6em of b] {$\nu$};
\end{feynman}\end{tikzpicture}
$=-i\tilde{\Delta}^{\mu \nu}(p)=-i\frac{g^{\mu \nu}-(1-\xi) \frac{p^{\mu} p^{\nu}}{p^{2}}}{p^{2}+i \epsilon}$\\
\\
\begin{tikzpicture}\begin{feynman}[small]
\diagram{a [dot] -- [boson, momentum=$p\text{,}\lambda$] b};
\vertex [above=0.6em of a] {$\nu$};
\end{feynman}\end{tikzpicture}
$=\epsilon^*_{\lambda}(p)^\nu$\\
\\
\begin{tikzpicture}\begin{feynman}[small]
\diagram{a -- [boson, momentum=$p\text{,}\lambda$] b [dot] };
\vertex [above=0.6em of b] {$\mu$};
\end{feynman}\end{tikzpicture}
$=\epsilon_{\lambda}(p)^\mu$

\end{minipage}
\begin{minipage}{0.5\textwidth}

\begin{tikzpicture}\begin{feynman}[small]
\vertex (a);\vertex [above left=of a] (b);\vertex [below left=of a] (c);\vertex [right=of a] (d);
\diagram{(b) -- [scalar, momentum=$p_1$, with arrow=0.5] (a) -- [photon] (d), (c) -- [scalar, momentum=$p_2$, with reversed arrow=0.5] (a)};
\vertex [above=0em of a] {$\;\;\;\;\;\;\mu$};
\end{feynman}\end{tikzpicture}
\raisebox{0.6cm}{$=-ie(p_1-p_2)^\mu$}\\
\\
\begin{tikzpicture}\begin{feynman}[small]
\vertex (v); \vertex [above left=of v] (1); \vertex [below left=of v] (2); \vertex [above right=of v] (3); \vertex [below right=of v] (4);
\diagram{(1) -- [scalar, with arrow=0.5] (v) -- [scalar, with arrow=0.5] (3), (2) -- [photon] (v) -- [photon] (4)};
\vertex [below=0em of v]{$\mu\;\;\;\;\;\;\;\;\;$};
\vertex [below=0em of v]{$\;\;\;\;\;\;\;\;\;\nu$};
\end{feynman}\end{tikzpicture}
\raisebox{0.6cm}{$=2ie^2g_{\mu\nu}$}\\
\\
\begin{tikzpicture}\begin{feynman}[small]
\vertex (v); \vertex [above left=of v] (1); \vertex [below left=of v] (2); \vertex [above right=of v] (3); \vertex [below right=of v] (4);
\diagram{(1) -- [scalar, with arrow=0.5] (v) -- [scalar, with reversed arrow=0.6] (3), (2) -- [scalar, with reversed arrow=0.5] (v) -- [scalar, with arrow=0.6] (4)};
\end{feynman}\end{tikzpicture}
\raisebox{0.6cm}{$=-i\lambda$}

\end{minipage}

\subsection{Spinor QED Feynman Rules} \label{QED}

\begin{minipage}{0.5\textwidth}

\begin{tikzpicture}\begin{feynman}[small]
\diagram{a [dot] -- [fermion, momentum=$p$] b [dot] };
\vertex [above=0.6em of a] {$\alpha$};
\vertex [above=0.6em of b] {$\beta$};
\end{feynman}\end{tikzpicture}
$=-i\tilde{S}_{\beta\alpha}(p)=i \frac{(\slashed{p}+m)_{\beta\alpha}}{p^{2}-m^{2}+i \epsilon}$\\
\\
\begin{tikzpicture}\begin{feynman}[small]
\diagram{a [dot] -- [boson, momentum=$p$] b [dot] };
\vertex [above=0.6em of a] {$\mu$};
\vertex [above=0.6em of b] {$\nu$};
\end{feynman}\end{tikzpicture}
$=-i\tilde{\Delta}^{\mu \nu}(p)=-i\frac{g^{\mu \nu}-(1-\xi) \frac{p^{\mu} p^{\nu}}{p^{2}}}{p^{2}+i \epsilon}$\\
\\
\begin{tikzpicture}\begin{feynman}[small]
\vertex (a); \vertex [right=of a] (b); \vertex [above right=of b] (c);\vertex [below right=of b] (d);
\diagram{(a) -- [photon] (b) -- [fermion] (c), (b) -- [anti fermion] (d)};
\vertex [above=0em of b] {$\mu\;\;\;\;\;\;\;$};
\vertex [above=0.6em of b] {$\;\;\alpha$};
\vertex [below=0.6em of b] {$\;\;\beta$};
\end{feynman}\end{tikzpicture}
\raisebox{0.8cm}{$=-ie\gamma^\mu_{\alpha\beta}$}

\end{minipage}
\begin{minipage}{0.5\textwidth}

\begin{tikzpicture}\begin{feynman}[small]
\diagram{a -- [fermion, momentum=$p\text{,}s$] b [dot] };
\vertex [above=0.6em of b] {$\alpha$};
\end{feynman}\end{tikzpicture}
$=u^s_\alpha(p)$
\quad\quad
\begin{tikzpicture}\begin{feynman}[small]
\diagram{a [dot] -- [fermion, momentum=$p\text{,}s$] b};
\vertex [above=0.6em of a] {$\beta$};
\end{feynman}\end{tikzpicture}
$=\bar{u}^s_\beta(p)$
\\
\begin{tikzpicture}\begin{feynman}[small]
\diagram{a -- [anti fermion, momentum=$p\text{,}s$] b [dot] };
\vertex [above=0.6em of b] {$\alpha$};
\end{feynman}\end{tikzpicture}
$=\bar{v}^s_\alpha(p)$
\quad\quad
\begin{tikzpicture}\begin{feynman}[small]
\diagram{a [dot] -- [anti fermion, momentum=$p\text{,}s$] b};
\vertex [above=0.6em of a] {$\beta$};
\end{feynman}\end{tikzpicture}
$=v^s_\beta(p)$
\\
\begin{tikzpicture}\begin{feynman}[small]
\diagram{a -- [boson, momentum=$p\text{,}\lambda$] b [dot] };
\vertex [above=0.6em of b] {$\mu$};
\end{feynman}\end{tikzpicture}
$=\epsilon_{\lambda}(p)^\mu$
\quad\;
\begin{tikzpicture}\begin{feynman}[small]
\diagram{a [dot] -- [boson, momentum=$p\text{,}\lambda$] b};
\vertex [above=0.6em of a] {$\nu$};
\end{feynman}\end{tikzpicture}
$=\epsilon^*_{\lambda}(p)^\nu$\\

\end{minipage}

\subsection{Gamma Matrix Technology} \label{Gamma}

\begin{minipage}{0.54\textwidth}
\begin{itemize}
\item $\{\gamma^\mu,\gamma^\nu\}=2g^{\mu\nu}\mathbbm{1}_4$
\item $g^{\mu\nu}g_{\mu\nu}=D$
\item $\slashed{A}:=\gamma^\mu A_\mu$
\item $(\gamma^\mu)^\dag=\gamma^0\gamma^\mu\gamma^0$
\item$\text{Tr}[\mbb{1_4}]=1$
\item$\text{Tr}[\text{any odd \# of $\gamma^\mu$'s}]=0$
\item $\operatorname{Tr}\left(\gamma^{\mu} \gamma^{\nu}\right)=4 g^{\mu \nu}$
\item $\operatorname{Tr}\left(\gamma^{\mu} \gamma^{\nu} \gamma^{\rho} \gamma^{\sigma}\right)=4\left(g^{\mu \nu} g^{\rho \sigma}-g^{\mu \rho} g^{\nu \sigma}+g^{\mu\sigma}g^{\nu\rho}\right)$
\end{itemize}
\end{minipage}
\begin{minipage}{0.46\textwidth}
\begin{itemize}
\item $\gamma^\mu\gamma_\mu=D\mbb{1}_4$
\item$\gamma^\mu\gamma^\nu\gamma_\mu=(2-D)\gamma^\nu$
\item$\gamma^\mu\gamma^\rho\gamma^\nu\gamma_\mu=4g^{\nu\rho}\mbb{1}_4-(4-D)\gamma^\rho \gamma^\nu$
\item$\gamma^\mu\gamma^\sigma\gamma^\rho\gamma^\nu\gamma_\mu=(4-D)\gamma^\sigma\gamma^\rho\gamma^\nu -2\gamma^\nu\gamma^\rho\gamma^\sigma$
\item $\gamma^5:=i\gamma^0\gamma^1\gamma^2\gamma^3$
\item$\{\gamma^5,\gamma^\mu\}=0$
\item$\gamma^{5\dag}=\gamma^5$
\item$\gamma^5\gamma^5=\mathbbm{1}$
\end{itemize}
\end{minipage}

\section{Axion-Photon Conversion Probability Derivation} \label{APP}

The Lagrangian for an interacting system of axions, photons, and EM currents is given by \begin{equation}\label{axion-photon-EM L} \mathcal{L}=\frac{1}{2} \partial_{\mu} a \partial^{\mu} a-\frac{1}{2} m_{\mathrm{a}}^{2} a^{2} -\frac{1}{4}F^{\mu\nu}F_{\mu\nu} -g_{a\gamma\gamma}a F^{\mu\nu}\tilde{F}_{\mu\nu}-J^\mu A_\mu\end{equation} where $-J^\mu A_\mu$ is the electromagnetic source term with the electric 4-current $J_\mu=(\rho,\mb{J})$. The electric and magnetic fields are then explicitly given by: \begin{equation}\label{EM fields} \mathbf{E}=-\boldsymbol{\nabla} A_{0}-\dot{\mathbf{A}},\quad \mathbf{B}=\boldsymbol{\nabla} \times \mathbf{A}.\end{equation}

Applying the Euler-Lagrange equations to (\ref{axion-photon-EM L}) yields the equations of motion: \begin{equation}\label{EoM} \begin{aligned} \partial_{\mu} F^{\mu \nu} &=J^{\nu}-g_{a \gamma} \tilde{F}^{\mu \nu} \partial_{\mu} a \\ \left(\partial_{\mu} \partial^{\mu}+m_{a}^{2}\right) a &=-\frac{g_{a \gamma}}{4} F_{\mu \nu} \tilde{F}^{\mu \nu}. \end{aligned}\end{equation} The first equation is a modification of Gauss' and Ampere's laws in the presence of axions and amounts to the addition of the extra current $J_{a}^{\nu} \equiv-g_{a \gamma} \partial_{\mu}\left(\widetilde{F}^{\mu \nu} a\right)=-g_{a \gamma} \widetilde{F}^{\mu \nu} \partial_{\mu} a$. The second equation is something totally new, the equation of motion of the axions. Writing these in terms of the electric and magnetic fields yields something rather familiar, yet different; the ALP-Maxwell equations: \begin{equation}\label{EoM2} \begin{aligned}
\boldsymbol{\nabla} \cdot \mathbf{E} &=\rho-g_{a \gamma} \mathbf{B} \cdot \boldsymbol{\nabla} a, \\
\boldsymbol{\nabla} \times \mathbf{B}-\dot{\mathbf{E}} &=\mathbf{J}+g_{a \gamma}(\mathbf{B} \dot{a}-\mathbf{E} \times \boldsymbol{\nabla} a) \\
\boldsymbol{\nabla} \cdot \mathbf{B} &=0 \\
\boldsymbol{\nabla} \times \mathbf{E}+\dot{\mathbf{B}} &=0 \\
\ddot{a}-\boldsymbol{\nabla}^{2} a+m_{a}^{2} a &=g_{a \gamma} \mathbf{E} \cdot \mathbf{B} .
\end{aligned}\end{equation}

When trying to detect axion photon conversion, a long optical cavity in a constant applied magnetic field is commonly used. Thus we may take $\mb{B}$ as a constant and the ALP-Maxwell equations (\ref{EoM2}) become linear and can be solved via a plane wave ansatz. Aligning our optical cavity along the $z$-direction, consider the ansatz for plane waves of frequency $\omega$ propagating along the cavity in the $z-$direction: \begin{equation} \label{ansatz} \begin{array}{l}
\left(\begin{array}{c}
a \\
i A_{\parallel} \\
i A_{\perp}
\end{array}\right)  \propto\left(\begin{array}{c}
\cos \vartheta \\
-\sin \vartheta \\
0
\end{array}\right) e^{-i\left(\omega t-k_{a}^{\prime} z\right)} \quad(\mathrm{ALP}-\operatorname{like}) \\
\left(\begin{array}{c}
a \\
i A_{\parallel} \\
i A_{\perp}
\end{array}\right)  \propto\left(\begin{array}{c}
\sin \vartheta \\
\cos \vartheta \\
0
\end{array}\right) e^{-i\left(\omega t-k_{\gamma}^{\prime} z\right)} \quad(\text { photon }-\operatorname{like}) \\
\left(\begin{array}{c}
a \\
i A_{\parallel} \\
i A_{\perp}
\end{array}\right)  \propto\left(\begin{array}{l}
0 \\
0 \\
1
\end{array}\right) e^{-i\left(\omega t-k_{\gamma} z\right)} \quad(\text { photon })
\end{array}\end{equation} where $\mb{A}:=\frac{\mb{E}}{i\omega}$, $A_\parallel$ has polarisation parallel to $z$, $A_\perp$ has polarisation perpendicular to both $z$ and $\mb{B}$, $k'_a$ and $k'_\gamma$ are modified wavenumbers, and $\vartheta$ is the \ti{mixing angle}. We seem to have arbitrarily snuck in the notion of ALP-photon mixing, but it is evidently present due to the interaction term in the Lagrangian (\ref{axion-photon-EM L}). In a background $\mb{B}$ field, neither photons nor ALPs correspond to freely propagating particles because the $g_{a\gamma\gamma}$ interaction acts as a non-diagonal mass term and quantum mechanically mixes ALPs with photons with polarisation along $\mb{B}$. The mixing angle and modified wavenumbers are given by: \begin{equation}\label{parameters} \begin{aligned}
\tan \vartheta &=\frac{2 g_{a \gamma} B_{e} \omega}{\Delta k^{2}+\sqrt{\left(\Delta k^{2}\right)^{2}+\left(2 g_{a \gamma} B_{e} \omega\right)^{2}}} \\
k_{\stackrel{a}{\gamma}}^{\prime 2} &=\frac{1}{2}\left(k_{\gamma}^{2}+k_{a}^{2} \mp \sqrt{\left(\Delta k^{2}\right)^{2}+\left(2 g_{a \gamma} B_{e} \omega\right)^{2}}\right)
\end{aligned}\end{equation} where $\Delta k^{2}:=k_{\gamma}^{2}-k_{a}^{2}$. Substituting this ansatz into (\ref{EoM2}) yields the solutions: \begin{equation}\label{axion sol} \begin{aligned}
E_{\|}(t, z) &=E_{\|, 0} e^{-i\left(\omega t-k_{\gamma}^{\prime} z\right)}\left[\cos ^{2} \vartheta+e^{-i q^{\prime} z} \sin ^{2} \vartheta\right] \\
a(t, z) &=\frac{E_{0}}{\omega} e^{-i\left(\omega t-k_{a}^{\prime} z\right)} \sin \vartheta \cos \vartheta\left[e^{i q^{\prime} z}-1\right]
\end{aligned}
\end{equation} where $q':=k'_\gamma-k_a'$. The component of the electromagnetic wave polarised perpendicular to the external $\mb{B}$ field will of course propagate as a standard electromagnetic wave with electric field \begin{equation} \label{E sol} E_{\perp}(t, z)=E_{ 0} e^{-i\left(\omega t-k_{\gamma} z\right)}.\end{equation}\\

The particle number densities of the photons and axions are given by \begin{equation}\label{particle number} \frac{|E_x|^2}{2\omega},\quad \frac{\omega|a|^2}{2}\end{equation} respectively. As we move away from $z=0$, the particle number densities of axions and photons change, but the total flux remains constant, which we can consider to be ALP-photon oscillation. The conversion probability at some distance $z$ is thus given by \begin{equation}\label{conversion prob}P(\gamma\to a)(z)=\frac{|a(z)|^2}{|A_x(z)|^2}=\left|\sin\vartheta\cos\vartheta\left[e^{i q^{\prime} z}-1\right]\right|^2=\sin^2(2\vartheta)\sin^2\left(\frac{q'z}{2}\right).\end{equation}

We can make two simplifying assumptions, small mixing $(\vartheta\ll1)$ and the relativistic limit $(\Delta k^2\approx2\omega q)$. We also note that, so long as we impose $\mb{E}(z=0)=0$, taking $\vartheta\to-\vartheta$ takes us from ALP-like
 to photon-like in (\ref{EoM2}). Since the transition probability is unchanged by $\vartheta\to-\vartheta$, photon to ALP oscillation probability upon propagating a distance $z$ is equal to the ALP to photon oscillation probability. Thus in our simplifying limits, the probability of conversion for an axion propagating along an optical cavity of length $L$ filled with a magnetic field of strength $B_e$ is: \begin{equation} P(\gamma\to a)=P(a\to\gamma)=\left(\frac{g_{a \gamma} B_{e}}{q}\right)^{2} \sin ^{2}\left(\frac{q L}{2}\right).\end{equation}

\end{appendices}

\end{singlespace}

\end{document}